\documentclass[onecolumn,twoside]{IEEEtran}
\usepackage{mathpazo}
\usepackage{times}
\usepackage{amsmath}
\usepackage{amsfonts}
\usepackage{latexsym}
\usepackage{amssymb}
\usepackage{mathrsfs}
\usepackage[dvipsnames]{xcolor}
\usepackage{upref}
\usepackage{theorem}
\usepackage{graphicx}
\usepackage{psfrag}
\usepackage{cite}

\usepackage{color}

\theoremstyle{plain}
\theorembodyfont{\normalfont\slshape}

\newtheorem{thm}{Theorem$\!$}
\newenvironment{theorem}
{\begin{thm}\hspace*{-1ex}{\bf.}}{\end{thm}}

\newtheorem{lem}[thm]{Lemma$\!$}
\newenvironment{lemma}{\begin{lem}\hspace*{-1ex}{\bf.}}{\end{lem}}

\newtheorem{alg}[thm]{Algorithm$\!$}

\newtheorem{prop}[thm]{Proposition$\!$}

\newtheorem{cor}[thm]{Corollary$\!$}
\newenvironment{corollary}{\begin{cor}\hspace*{-1ex}{\bf.}}{\end{cor}}

\newtheorem{defn}[thm]{Definition$\!$}
\newenvironment{definition}{\begin{defn}\hspace*{-1ex}{\bf.}}{\end{defn}}

\newtheorem{xmpl}[thm]{Example$\!$}
\newenvironment{example}{\begin{xmpl}\hspace*{-1ex}{\bf.}}{\hfill$\Box$\end{xmpl}}

\newtheorem{cnstr}{Construction$\!$}

\newcounter{enumrom}
\renewcommand{\theenumrom}{(\roman{enumrom})}

\makeatletter
\renewcommand{\@endtheorem}{\endtrivlist}
\makeatother

\makeatletter
\renewcommand{\thefigure}{{\@arabic\c@figure}}
\renewcommand{\fnum@figure}{{\bf Figure\,\thefigure}}
\makeatother

\newcommand{\pf}{{\bf Proof: }}

\newcommand{\uv}{\mbox{$\underline{v}$}}

\newcommand{\be}[1]{\begin{equation}\label{#1}}
\newcommand{\ee}{\end{equation}}

\renewcommand{\leq}{\leqslant}

\renewcommand{\geq}{\geqslant}

\newcommand{\Cref}[1]{Co\-ro\-lla\-ry\,\ref{#1}}

\newcommand{\C}{\mbox{${\cal C}$}}
\newcommand{\R}{\mbox{${\cal R}$}}

\newcommand{\lan}{\mbox{$\langle$}}
\newcommand{\ran}{\mbox{$\rangle$}}
\newcommand{\qed}{\hfill$\Box$\\[1ex]}

\newcommand{\uc}{\mbox{$\underline{c}$}}

\newcommand{\al}{\alpha}

\newcommand{\xor}{\oplus}

\newcommand{\rc}{\textcolor{red}}
\newcommand{\bc}{\textcolor{blue}}

\newcommand{\ue}{\underline{e}}
\newcommand{\uz}{\underline{z}}

\newcommand{\eq}{\mbox{$\,=\,$}}

\outer\def\proclaim #1. #2\par{\medbreak
 \noindent{\bf#1.\enspace}{\sl#2\par}%
 \ifdim\lastskip<\medskipamount \removelastskip\penalty55\medskip\fi}

\begin{document}


\title{\LARGE\bf Array Codes with Local Properties}

\author{\large
Mario~Blaum and Steven R. Hetzler\\
IBM Research Division\\
Almaden Research Center\\
 San Jose, CA 95120, USA \\
}


\maketitle

\begin{abstract}
In general, array codes consist of $m\times n$ arrays and in many
cases, the arrays satisfy parity constraints along lines of different
slopes (generally with a toroidal topology). Such codes are useful
for RAID type of architectures, since they allow to replace finite
field operations by XORs. We present expansions to traditional array
codes of this type, like Blaum-Roth (BR) and extended EVENODD codes,
by adding parity on columns. This vertical parity allows for recovery
of one or more symbols in a column locally, i.e., by using the
remaining symbols in the column without invoking the rest of the array.
Properties and applications of the new codes are discussed, in
particular to Locally Recoverable (LRC) codes.
\end{abstract}


\section{Introduction}
\label{Introduction}
Throughout this paper $p$ will denote a prime number and $q$ a power
of 2, i.e., $q\eq 2^b$. We will consider finite fields $GF(q)$. The
results can be extended to other prime powers, but for simplicity, we
consider finite fields of characteristic 2 only.

Given $p$ and an integer $m$, let $\lan
m\ran_p$ be the unique number $j$, $0\leq j\leq p-1$ such that $j\equiv m\;(\bmod\;p)$.
Unless there is confusion, we denote $\lan m\ran_p$ as
$\lan m\ran$. We will consider
$p\times p$ arrays with entries $c_{u,v}\in GF(q)$, $0\leq u,v\leq
p-1$. We next define formally a line in an array.

\begin{definition}
\label{def0}
{\em
Given a $p\times p$ array with entries $c_{u,v}\in GF(q)$, $0\leq
u,v\leq p-1$, a line of slope $i$, $0\leq i\leq p-1$, through
entry $c_{u_0,0}$, $0\leq u_0\leq p-1$, is
the set of $p$ entries $\{c_{\lan u_0-iv\ran,v}\,:\,0\leq v\leq p-1\}$. A line
of slope $\infty$ (vertical line) through entry
$(0,v_0)$, $0\leq v_0\leq p-1$, is
the set of $p$ entries $\{c_{u,v_0}\,:\,0\leq u\leq p-1\}$.
 \qed
}
\end{definition}

Next we give the definition of Blaum-Roth (BR) codes~\cite{br}:

\begin{definition}
\label{defBR0}
{\em
A BR code with $r$
parity columns, denoted $BR(p,r,q)$,
consists of all possible $(p-1)\times p$ arrays over $GF(q)$ such
that, when a zero row is appended to an array in the code, each line of slope
$i$ for $0\leq i\leq r-1$ as given by Definition~\ref{def0} has even
parity.
 \qed
}
\end{definition}

\begin{example}
\label{ex00}
{\em
The first 4 rows of the
following $5\times 5$ array are in $BR(5,3,2)$:

$$
\begin{array}{ccc}
\begin{array}{|c|c|c|c|c|}
\hline
1& 0& 0& 0& 1\\\hline
\rc{1}& \rc{1}& \rc{1}& \rc{0}& \rc{1}\\\hline
0& 1& 0& 0& 1\\\hline
0& 1& 0& 0& 1\\\hline\hline
0& 0& 0& 0& 0\\
\hline
\end{array}
&
\begin{array}{|c|c|c|c|c|}
\hline
1& \rc{0}& 0& 0& 1\\\hline
\rc{1}& 1& 1& 0& 1\\\hline
0& 1& 0& 0& \rc{1}\\\hline
0& 1& 0& \rc{0}& 1\\\hline\hline
0& 0& \rc{0}& 0& 0\\
\hline
\end{array}
&
\begin{array}{|c|c|c|c|c|}
\hline
1& 0& 0& \rc{0}& 1\\\hline
\rc{1}& 1& 1& 0& 1\\\hline
0& 1& \rc{0}& 0& 1\\\hline
0& 1& 0& 0& \rc{1}\\\hline\hline
0& \rc{0}& 0& 0& 0\\
\hline
\end{array}
\end{array}
$$

We can see that every
horizontal line (i.e., line of slope 0), every line of slope 1 and
every line of slope 2 have even parity (we mark in red the lines of
slope 0, 1 and 2 through entry $c_{1,0}$ respectively).
}
\end{example}

The following algebraic definition~\cite{br} of a $BR(p,r,q)$ code
is equivalent to Definition~\ref{def0}. This definition is convenient
for decoding purposes.

\begin{definition}
\label{defBR}
{\em
A $BR(p,r,q)$ code is the code over the ring of polynomials modulo
$M_{p,q}(x)\eq 1\xor x\xor x^2+\cdots \xor x^{p-1}$ with coefficients
in $GF(q)$ whose parity-check matrix is given
by the $r\times p$ Reed-Solomon type of matrix

\begin{eqnarray}
\label{Hpr}
H_{p,r}&=&\left(
\begin{array}{ccccc}
1&1&1&\ldots &1\\
1&\al &\al^2&\ldots &\al^{p-1}\\
1&\al^2 &\al^4&\ldots &\al^{2(p-1)}\\
\vdots &\vdots &\vdots &\ddots &\vdots\\
1&\al^{r-1} &\al^{2(r-1)}&\ldots &\al^{(r-1)(p-1)}\\
\end{array}
\right),
\end{eqnarray}
where $M_{p,q}(\al)\eq 0$ and $\al\neq 1$. 
 \qed
}
\end{definition}

From Definition~\ref{defBR}, each codeword in $BR(p,r,q)$ can be
considered as a $(p-1)\times p$ array such that each column
represents an element in the ring of polynomials modulo
$M_{p,q}(x)$. It can be verified that such an array
satisfies Definition~\ref{defBR0}~\cite{br}.
At this point the field $GF(q)$ is not significative. In fact, since
$q\eq 2^b$, Definition~\ref{defBR} uses $b$ codes $BR(p,r,2)$ in
parallel, so studying $BR(p,r,q)$ codes is equivalent to studying
$BR(p,r,2)$ codes. This will change with the
expansion of $BR(p,r,q)$ codes to be presented next, which incorporates
vertical parities in each column of the array. 


\begin{definition}
\label{defGEBR}
{\em
Let 
$g(x)$ be a polynomial with coefficients in 
$GF(q)$ such
that $g(x)$ divides $1\xor x^p$ and $\gcd(1\xor x,g(x))\eq 1$. Denote
by $\C(p,g(x)(1\xor x),q,d)$ the cyclic code
of length $p$ over $GF(q)$ whose generator
polynomial is $g(x)(1\xor x)$ having minimum distance $d$.
Let $\R_p(q)$ be the ring of polynomials modulo
$1\xor x^p$ with coefficients in $GF(q)$,
and let $\al\in\R_p(q)$ such that $\al^p\eq 1$ and $\al\neq 1$. Then an Expanded
Blaum-Roth code $EBR(p,r,q,g(x))$ is the $[p,p-r]$ code with
coefficients in $\C(p,g(x)(1\xor x),q,d)$ 
whose parity-check matrix is given by~(\ref{Hpr}). \qed
}
\end{definition}

Notice that $\C(p,g(x)(1\xor x),q,d)$ is an ideal in $\R_p(q)$.
The definition of an $EBR(p,r,q,g(x))$ code looks similar to the
one of a $BR(p,r,q)$
code. They both share the parity-check matrix~(\ref{Hpr}), but while
the entries of a codeword in a $BR(p,r,q)$ code are in the ring of polynomials modulo
$M_{p,q}(x)$, 
the entries of a codeword in an $EBR(p,r,q,g(x))$ code are in the
ideal $\C(p,g(x)(1\xor x),q,d)$. 
Thus, a codeword in an $EBR(p,r,q,g(x))$ code can be represented by
a $p\times p$ array, where each column is in the cyclic code
$\C(p,g(x)(1\xor x),q,d)$. We also observe that 
$\al^i$ represents a rotation $i$ times. Hence, multiplying by $\al^i$ does
not involve finite field arithmetic or XOR operations. Let us point
out that codes over
the ring $\R_m(q)$, $m$ an integer (not necessarily prime), have been
used in literature, for example, in~\cite{hhsl} to provide a
unification between EVENODD and RDP codes, in~\cite{hscl} in the
context of Regenerating Codes, and in~\cite{hscl2} to give
efficient encoding and decoding algorithms for a family of MDS array
codes.

From Definition~\ref{defGEBR} we obtain a geometrical description of the
codes. As stated above, 
the codewords in an $EBR(p,r,q,g(x))$ code can be represented as
$p\times p$ arrays over 
$GF(q)$ such that
each column (i.e., line of slope $\infty$) is in the code\\
$\C(p,g(x)(1\xor x),q,d)$, and in addition, any line of slope $j$ for
$0\leq j\leq r-1$ has even parity (recall that the arrays in a
$BR(p,r,q)$ code are $(p-1)\times p$ arrays). 

An important special case is given by code $EBR(p,r,2,1)$, as described
in~\cite{bdh}, i.e., $g(x)\eq 1$ and each column in an array has even
parity, as illustrated in the next example.

\begin{example}
\label{ex13}
{\em
The following is an array in
$EBR(5,3,2,1)$:

$$
\begin{array}{cccc}
\begin{array}{|c|c|c|c|c|}
\hline
\rc{1}& 0& 0& 1& 0\\\hline
\rc{1}& 1& 1& 0& 1\\\hline
\rc{0}& 1& 1& 0& 0\\\hline
\rc{0}& 1& 1& 0& 0\\\hline
\rc{0}& 1& 1& 1& 1\\
\hline
\end{array}
&
\begin{array}{|c|c|c|c|c|}
\hline
1& 0& 0& 1& 0\\\hline
\rc{1}& \rc{1}& \rc{1}& \rc{0}& \rc{1}\\\hline
0& 1& 1& 0& 0\\\hline
0& 1& 1& 0& 0\\\hline
0& 1& 1& 1& 1\\
\hline
\end{array}
&
\begin{array}{|c|c|c|c|c|}
\hline
1& \rc{0}& 0& 1& 0\\\hline
\rc{1}& 1& 1& 0& 1\\\hline
0& 1& 1& 0& \rc{0}\\\hline
0& 1& 1& \rc{0}& 0\\\hline
0& 1& \rc{1}& 1& 1\\
\hline
\end{array}
&
\begin{array}{|c|c|c|c|c|}
\hline
1& 0& 0& \rc{1}& 0\\\hline
\rc{1}& 1& 1& 0& 1\\\hline
0& 1& \rc{1}& 0& 0\\\hline
0& 1& 1& 0& \rc{0}\\\hline
0& \rc{1}& 1& 1& 1\\
\hline
\end{array}
\end{array}
$$

We can see that each column has even parity (i.e., each column is in
$\C(p,1\xor x,2,2)$), as well as each line of slope 0, 1 and 2
(illustrated in red for the lines through entry $c_{1,0}$).
}
\end{example}

\begin{example}
\label{ex8}
{\em
The code $EBR(7,3,2,1\xor x\xor x^3)$ consists of
all the $7\times 7$ arrays having even parity on lines of slopes 0, 1
and 2 and whose columns are in the code $\C(7,(1\xor x\xor
x^3)(1\xor x),2,4)$, which is the subcode of the $[7,4,3]$ cyclic
Hamming code generated by $1\xor x\xor x^3$ whose codewords have even
weight~\cite{ms}.

The following are two arrays in $EBR(7,3,2,1\xor x\xor x^3)$:

$$
\begin{array}{cc}
\begin{array}{|c|c|c|c||c|c|c|}
\hline
1&0&1&0&1&0&1\\
\hline
1&1&1&0&0&0&1\\
\hline
0&1&1&0&0&1&1\\
\hline\hline
0&1&0&0&1&0&0\\
\hline
1&0&0&0&0&1&0\\
\hline
0&0&1&0&1&1&1\\
\hline
1&1&0&0&1&1&0\\
\hline
\end{array}
&
\begin{array}{|c|c|c|c||c|c|c|}
\hline
0&0&0&0&0&1&1\\
\hline
0&0&0&1&1&0&0\\
\hline
0&0&0&1&0&1&0\\
\hline\hline
0&0&0&1&1&1&1\\
\hline
0&0&0&0&1&1&0\\
\hline
0&0&0&0&1&0&1\\
\hline
0&0&0&1&0&0&1\\
\hline
\end{array}
\end{array}
$$

We will see in Theorem~\ref{theo1} that, like a BR code, an EBR code
has minimum distance $r+1$ on columns. As a consequence,
the minimum Hamming distance of this code when considered as a
code over $GF(2)$ is $D\eq 16$. In effect, taking a non-zero array in
the code, each
non-zero column has weight at least four and there are at least four non-zero
columns, so $D\geq 16$, while the array in the right
above has weight exactly 16. As a comparison, the product code
consisting of the product of 
$\C(7,(1\xor x\xor x^3)(1\xor x),2,4)$ with the $[7,4,3]$ Hamming
code generated by $1\xor x\xor x^3$
has the same rate as\\ $EBR(7,3,2,1\xor x\xor x^3)$ but minimum Hamming
distance 12.
}
\end{example}

\begin{example}
\label{ex9}
{\em
Let $p\eq q-1$ be a prime number (in fact, a Mersenne prime), 
$q\eq 2^b$,  $\beta$ a primitive
element in $GF(q)$,\\ $g_{D-2}(x)\eq\prod_{i=1}^{D-2}(\beta^i\xor x)$ and consider the
RS code $\C(p,g_{D-2}(x)(1\xor x),q,D)$~\cite{ms}. Then the
$EBR(p,r,q,g_{D-2}(x))$ code as given by Definition~\ref{defGEBR} consists of
all the $p\times p$ arrays over $GF(q)$ having even parity on lines of slopes $j$,
$0\leq j\leq r-1$, and whose columns are in the RS code
$\C(p,g_{D-2}(x)(1\xor x),q,D)$.

For example, we can take $p\eq 7\eq 8-1$, $\beta$ a primitive element
in $GF(8)$ and $g_1(x)\eq\beta \xor x$.
Then, according to Definition~\ref{defGEBR}, the code
$EBR(7,3,8,g_{1}(x))$ consists of all the $7\times 7$ arrays over
$GF(8)$ having even parity on lines of slope 0, 1 and 2, and whose
columns are in the $[7,5]$ RS code $\C(7,(1\xor x)(\beta \xor x),8,3)$.

More concretely, if $\beta$ is a zero of the primitive polynomial
$1\xor x\xor x^3$, the reader can verify that the following is an array in $EBR(7,3,8,g_{1}(x))$:

$$
\begin{array}{|c|c|c|c||c|c|c|}
\hline
0 & \beta ^4 & 1& 0 & 1& \beta & \beta ^2\\\hline
\beta ^5 & \beta ^6 & \beta ^5 & \beta ^4 & \beta & \beta ^6 & \beta ^2 \\\hline
\beta ^5 & \beta ^5 & \beta ^4 & 0 & \beta & \beta ^3 & \beta ^5 \\\hline
\beta ^2 & \beta ^4 & 0 & \beta ^2 & 0 & \beta ^4 & 0 \\\hline
\beta ^2 & \beta ^4 & \beta & \beta ^3 & \beta ^4 & 1& \beta ^2 \\\hline\hline
\beta ^3 & \beta ^4 & \beta ^4 & \beta & 1& \beta ^4 & \beta ^4 \\\hline
\beta ^3 & \beta & \beta ^2 & \beta ^3 & \beta ^4 & \beta ^6 & \beta ^6 \\\hline
\end{array}
$$

We may assume that the first 5 symbols in the first 4 columns are a
$5\times 4$ data array, while the remaining symbols are parity symbols.
}
\end{example}

The next lemma establishes the connection between $BR(p,r,q)$ and
$EBR(p,r,q,1)$ codes.

\begin{lemma}
\label{l0}
{\em There is a 1-1 relationship between $BR(p,r,q)$ and
$EBR(p,r,q,1)$ codes preserving the (column) weight of each array in the code.
}
\end{lemma}
\pf
Consider an array in $EBR(p,r,q,1)$ of (column) weight $w$. Denote the array as
$C\eq (\uc_0,\uc_1,\ldots,\uc_{p-1})$, where\\ $\uc_j\eq
(c_{0,j},c_{1,j},\ldots,c_{p-1,j})$ is a (column) vector of length
$p$ for $0\leq j\leq p-1$. For each $\uc_j$, define a (column)
vector of length $p-1$ $\hat{\uc}_j\eq
(\hat{c}_{0,j},\hat{c}_{1,j},\ldots,\hat{c}_{p-2,j})$ such that
$\hat{c}_{i,j}\eq c_{i,j}\xor c_{p-1,j}$ for $0\leq i\leq p-2$.
Then, consider the transformation from $EBR(p,r,q,1)$ to $BR(p,r,q)$
given by

\begin{eqnarray}
\label{eq10}
C\;\eq\;(\uc_0,\uc_1,\ldots,\uc_{p-1})&\rightarrow &(\hat{\uc}_0,\hat{\uc}_1,\ldots,\hat{\uc}_{p-1})\;\eq\;\hat{C}.
\end{eqnarray}

First we need to prove that $\hat{C}$
in~(\ref{eq10}) is
in $BR(p,r,q)$. Since an array in $EBR(p,r,q,1)$ (resp. $BR(p,r,q)$)
consists of $q$ independent arrays in $EBR(p,r,2,1)$ (resp.
$BR(p,r,2)$), without loss of generality, we assume $q\eq 2$. By
Definition~\ref{defBR0}, $\hat{C}\in BR(p,r,2)$
if and only if every line of slope $s$ in the $p\times p$
array consisting of $\hat{C}$ with a zero row appended at the bottom
has even parity for $0\leq s\leq r-1$. Notice that, by~(\ref{eq10}), such an array
is equal to $C\xor W$, where $W$ is a $p\times p$ array such that
column $j$ of $W$ is an all-zero vector if $c_{p-1,j}\eq 0$,
otherwise it is an all-one vector. Since, in particular the weight of
row $p-1$ in $C$ is even, the number of all-one columns in $W$ is
even. Hence, any line of slope $s$, $0\leq s\leq r-1$, has even
parity in $C\xor W$ and $\hat{C}\in BR(p,r,2)$.

Next we have to show that $C$ and $\hat{C}$ have the same (column)
weight. This is true because $\uc_j$ is non-zero if and only if
$\hat{\uc}_j$ is non-zero for $0\leq j\leq p-1$. In effect, if
$\uc_j$ is non-zero and $c_{p-1,j}\eq 0$, then
$c_{i,j}\eq\hat{c}_{i,j}$ for $0\leq i\leq p-2$ and $\hat{\uc}_j$ is
non-zero. If $\uc_j$ is non-zero and $c_{p-1,j}\eq 1$, then the number of
1s in $\uc_{i,j}$ for $0\leq i\leq p-2$ is odd, as well as the number
of zeros. Hence, the number of 1s in $\hat{c}_{i,j}\eq 1\xor c_{i,j}$
for $0\leq i\leq p-2$ is odd, thus
$\hat{\uc}_j$ has odd weight and it cannot be a zero vector.
\qed

\begin{cor}
\label{cor0}
{\em
The code $EBR(p,r,q,g(x))$ over $\C(p,g(x)(1\xor x),q,d)$ given by
Definition~\ref{defGEBR} is MDS.
}
\end{cor}
\pf
Let $C$ be an array in $EBR(p,r,q,g(x))$ of (column) weight $w$. Applying
transformation~(\ref{eq10}) to $C$, we obtain\\ $\hat{C}\in BR(p,r,q)$.
Since, by Lemma~\ref{l0}, $\hat{C}$ has also (column) weight $w$, and
since $BR(p,r,q)$ is MDS~\cite{br}, then $w\geq r+1$ and also
$EBR(p,r,q,g(x))$ is MDS. 
\qed

\begin{example}
\label{ex15}
{\em
Consider the array in $EBR(5,3,2,1)$ given in Example~\ref{ex13}.
Then the following is the transformation of this array into an array
in $BR(5,3,2)$ as given by~(\ref{eq10}), where we add a row of zeros
to the $4\times 5$ array in $BR(5,3,2)$:

\begin{eqnarray*}
\begin{array}{|c|c|c|c|c|}
\hline
1& 0& 0& 1& 0\\\hline
1& 1& 1& 0& 1\\\hline
0& 1& 1& 0& 0\\\hline
0& 1& 1& 0& 0\\\hline
0& 1& 1& 1& 1\\
\hline
\end{array}
&\rightarrow &
\begin{array}{|c|c|c|c|c|}
\hline
1& 1& 1& 0& 1\\\hline
1& 0& 0& 1& 0\\\hline
0& 0& 0& 1& 1\\\hline
0& 0& 0& 1& 1\\\hline\hline
0& 0& 0& 0& 0\\\hline
\end{array}.
\end{eqnarray*}

We can see that the parity along all the lines of slope 0, 1 and 2 is
preserved.
}
\end{example}

Since it is well known how to encode and decode $BR(p,r,q)$
codes~\cite{br,hhsl}, the same can be done for $EBR(p,r,q,g(x))$ codes. In
effect, the first step in the decoding consists of recovering
up to $d-1$ erasures in each column (i.e., locally)
whenever possible using the cyclic code
$\C(p,g(x)(1\xor x),q,d)$ (since the code is cyclic, also a burst of
up to $1+\deg(g)$ erasures can be recovered, as is the case in the encoding).
Once this is done, the array in
$EBR(p,r,q,g(x))$ is transformed into an array in $BR(p,r,q)$ using
transformation~(\ref{eq10}). This array is decoded in $BR(p,r,q)$ using,
for example, the method in~\cite{br} or
in~\cite{hhsl}. Once the decoding is complete,
the inverse transformation is applied to obtain the original array in
$EBR(p,r,q,g(x))$. More efficient encoding and decoding methods,
though, will be presented in Section~\ref{EBR}.

Quite often, it is desirable that the parity columns in an array code
are independent, like in~\cite{bbbm,bbv,br2}, 
since this property allows for a
minimization of the number of parity updates when a data symbol is
updated. Next we will do the same for
array codes with local properties. We start with the definition of
Independent Parity (IP) codes~\cite{bbbm,bbv}.

\begin{definition}
\label{defIP0}
{\em
An IP code with $r$
parity columns, denoted $IP(p,r,q)$,
consists of all possible $(p-1)\times (p+r)$ arrays over $GF(q)$ such
that, when a zero row is appended to the array, for each $s$, $0\leq
s\leq r-1$, the $p$ sets
$$\{(u,v)\,:\,u+sv\eq i\}\cup\{(i,p+s)\}$$ for $0\leq i\leq p-1$ have
all the same parity, either even or odd. \qed
}
\end{definition}

$IP(p,r,q)$ codes are also known as
Generalized EVENODD codes~\cite{bbbmv} and Blaum--Bruck--Vardy
codes~\cite{hsl} in literature.

\begin{example}
\label{ex40}
{\em
The $4\times 8$ array consisting of the first 4 rows of the
following $5\times 8$ array is in $IP(5,3,2)$:

$$
\begin{array}{ccc}
\begin{array}{|c|c|c|c|c||c|c|c|}
\hline
1&0&0&1&1&1&1&1\\
\hline
\rc{0}&\rc{1}&\rc{0}&\rc{1}&\rc{1}&\rc{1}&1&0\\
\hline
0&0&0&0&1&1&1&1\\
\hline
1&1&0&1&1&0&1&1\\
\hline\hline
0&0&0&0&0&0&0&0\\
\hline
\end{array}
\hspace{-.6mm}
&
\hspace{-.6mm}
\begin{array}{|c|c|c|c|c||c|c|c|}
\hline
1&\rc{0}&0&1&1&1&1&1\\
\hline
\rc{0}&1&0&1&1&1&\rc{1}&0\\
\hline
0&0&0&0&\rc{1}&1&1&1\\
\hline
1&1&0&\rc{1}&1&0&1&1\\
\hline\hline
0&0&\rc{0}&0&0&0&0&0\\
\hline
\end{array}
\hspace{-.6mm}
&
\hspace{-.6mm}
\begin{array}{|c|c|c|c|c||c|c|c|}
\hline
1&0&0&\rc{1}&1&1&1&1\\
\hline
\rc{0}&1&0&1&1&1&1&\rc{0}\\
\hline
0&0&\rc{0}&0&1&1&1&1\\
\hline
1&1&0&1&\rc{1}&0&1&1\\
\hline\hline
0&\rc{0}&0&0&0&0&0&0\\
\hline
\end{array}
\end{array}
$$

We can see that given $s$, $0\leq s\leq 2$, all the lines of slope $s$ together
with the corresponding independent parity in column $5+s$
(illustrated in red for the lines through entry $c_{1,0}$) have the
same parity: even for lines of slope 0, odd for lines of slope 1 and
even for lines of slope 2. 
}
\end{example}

Similarly to $BR(p,r,q)$ codes, there is an equivalent
algebraic description of $IP(p,r,q)$ codes~\cite{bbbm,bbv}. Explicitly:

\begin{definition}
\label{defIP}
{\em
An $IP(p,r,q)$ code is the code over the ring of polynomials modulo
$M_{p,q}(x)$
whose parity-check matrix is given
by the $r\times (p+r)$ matrix

\begin{eqnarray}
\label{HEOpr}
\tilde{H}_{p,r}&=&\left(
\begin{array}{ccccccccccc}
1&1&1&\ldots &1&1&0&0&0&\ldots &0\\
1&\al &\al^2&\ldots &\al^{p-1}&0&1&0&0&\ldots &0\\
1&\al^2 &\al^4&\ldots &\al^{2(p-1)}&0&0&1&0&\ldots &0\\
\vdots &\vdots &\vdots &\ddots &\vdots&\vdots&\vdots&\vdots&\ddots&\vdots&\vdots\\
1&\al^{r-1} &\al^{2(r-1)}&\ldots &\al^{(r-1)(p-1)}&0&0&0&0&\ldots &1\\
\end{array}
\right),
\end{eqnarray}
where $M_{p,q}(\al)\eq 0$ and $\al\neq 1$.
 \qed
}
\end{definition}

By Definition~\ref{defIP}, each array in $IP(p,r,q)$ can be
considered as a $(p-1)\times (p+r)$ array such that each column
represents an element in the ring of polynomials modulo
$M_{p,q}(x)$.

Next we define Expanded Independent Parity (EIP) codes:

\begin{definition}
\label{defEEO}
{\em
Consider the cyclic code $\C(p,g(x)(1\xor x),q,d)$ and
$\al\in\R_p(q)$ as in Definition~\ref{defGEBR}.
Then an EIP code\\ $EIP(p,r,q,g(x))$ is the $[p+r,p]$ code
over $\C(p,g(x)(1\xor x),q,d)$ 
whose parity-check matrix $\tilde{H}_{p,r}$ is given by~(\ref{HEOpr}). \qed
}
\end{definition}

Contrary to $IP(p,r,q)$ codes, the parities in $EIP(p,r,q,g(x))$ codes are
always even.

The next example illustrates Definition~\ref{defEEO}.

\begin{example}
\label{ex14}
{\em
The following is an array in $EIP(5,3,2,1)$ (the data, correspoding to
the first 4 rows and the first 5 columns, is the same as in the array
given in Example~\ref{ex40}): 

$$
\begin{array}{cccc}
\begin{array}{|c|c|c|c|c||c|c|c|}
\hline
\rc{1}&0&0&1&1&1&0&0\\
\hline
\rc{0}&1&0&1&1&1&0&0\\
\hline
\rc{0}&0&0&0&1&1&1&1\\
\hline
\rc{1}&1&0&1&1&0&0&1\\
\hline\hline
\rc{0}&0&0&1&0&1&1&0\\
\hline
\end{array}
\hspace{-.6mm}
&
\hspace{-.6mm}
\begin{array}{|c|c|c|c|c||c|c|c|}
\hline
1&0&0&1&1&1&0&0\\
\hline
\rc{0}&\rc{1}&\rc{0}&\rc{1}&\rc{1}&\rc{1}&0&0\\
\hline
0&0&0&0&1&1&1&1\\
\hline
1&1&0&1&1&0&0&1\\
\hline\hline
0&0&0&1&0&1&1&0\\
\hline
\end{array}
\hspace{-.6mm}
&
\hspace{-.6mm}
\begin{array}{|c|c|c|c|c||c|c|c|}
\hline
1&\rc{0}&0&1&1&1&0&0\\
\hline
\rc{0}&1&0&1&1&1&\rc{0}&0\\
\hline
0&0&0&0&\rc{1}&1&1&1\\
\hline
1&1&0&\rc{1}&1&0&0&1\\
\hline\hline
0&0&\rc{0}&1&0&1&1&0\\
\hline
\end{array}
\hspace{-.6mm}
&
\hspace{-.6mm}
\begin{array}{|c|c|c|c|c||c|c|c|}
\hline
1&0&0&\rc{1}&1&1&0&0\\
\hline
\rc{0}&1&0&1&1&1&0&\rc{0}\\
\hline
0&0&\rc{0}&0&1&1&1&1\\
\hline
1&1&0&1&\rc{1}&0&0&1\\
\hline\hline
0&\rc{0}&0&1&0&1&1&0\\
\hline
\end{array}
\end{array}
$$

Each column has even parity (i.e., each column is in
$\C(p,1\xor x,2,2)$), as well as each line of slope 0, 1 and 2 together
with the corresponding independent parity
(illustrated in red for the lines through entry $c_{1,0}$).
}
\end{example}

The next lemma, similarly to Lemma~\ref{l0}, establishes the
connection between the codes $IP(p,r,q)$ and $EIP(p,r,q,1)$:

\begin{lemma}
\label{l7}
{\em There is a 1-1 relationship between $IP(p,r,q)$ and
$EIP(p,r,q,1)$ preserving the (column) weight of each array in the code.
}
\end{lemma}

\pf
Proceeding as in Lemma~\ref{l0}, denote an array $C\in EIP(p,r,q,1))$
as $C\eq (\uc_0,\uc_1,\ldots,\uc_{p+r-1})$, where each $\uc_j$ is a
(column) vector of length
$p$ for $0\leq j\leq p+r-1$ and $\hat{\uc}_j$ is defined as in the
proof of Lemma~\ref{l0}.
Then, consider the transformation from $EIP(p,r,q,1)$ to $IP(p,r,q)$
given by

\begin{eqnarray}
\label{eq100}
C\;\eq\;(\uc_0,\uc_1,\ldots,\uc_{p+r-1})&\rightarrow
&(\hat{\uc}_0,\hat{\uc}_1,\ldots,\hat{\uc}_{p+r-1})\;\eq\;\hat{C}.
\end{eqnarray}

We have to prove that $\hat{C}$ in~(\ref{eq100}) is
in $IP(p,r,q)$. As in Lemma~\ref{l0}, without loss of generality we
may assume that $q\eq 2$. Consider the $p\times (p+r)$ array
consisting of $\hat{C}$ with a zero-row appended. By
Definition~\ref{defIP0} of $IP(p,r,2)$, we have to prove 
that in such $p\times (p+r)$ array, the lines of slope
$s$ in the first $p$ columns of the array, $0\leq s\leq r-1$,
starting in entry $\hat{c}_{i,0}$, $0\leq i\leq p-1$, together with
entry $\hat{c}_{i,p+s}$, have all the same parity, either even or odd.
As in Lemma~\ref{l0}, by~(\ref{eq100}), such an array
is equal to $C\xor W$, where $W$ is a $p\times (p+r)$ array such that
column $j$ of $W$ is an all-zero vector if $c_{p-1,j}\eq 0$,
otherwise it is an all-one vector.
Consider vector $\uv_s\eq (c_{p-1,0},c_{p-1,1},\ldots,c_{p-1,p-1},c_{p-1,p+s})$,
where $0\leq s\leq r-1$, and let $W_s$ be the $p\times (p+1)$ matrix
consisting of columns $0,1,\ldots,p-1,p+s$ of $W$. If the weight of
$\uv_s$ is even, then the number of all-one columns in $W_s$ is
even and any line of slope
$s$ in the first $p$ columns of the array
through entry $\hat{c}_{i,0}$, $0\leq i\leq p-2$ (as given by
Definition~\ref{def0}), together with 
entry $\hat{c}_{i,p+s}$, has even parity. Otherwise, all such lines together with
entry $\hat{c}_{i,p+s}$ have odd parity.

Regarding the weight preservation, the argument is the same as in Lemma~\ref{l0}.
\qed

\begin{example}
\label{ex16}
{\em
Consider the array in $EIP(5,3,2,1)$ given in Example~\ref{ex14}. The
transformation from $EIP(5,3,2,1)$ to $IP(5,3,2)$ given
by~(\ref{eq100}) is (by appending a row of zeros to the $4\times 8$
array in $IP(5,3,2,1)$)

\begin{eqnarray*}
\begin{array}{|c|c|c|c|c||c|c|c|}
\hline
1&0&0&1&1&1&0&0\\
\hline
0&1&0&1&1&1&0&0\\
\hline
0&0&0&0&1&1&1&1\\
\hline
1&1&0&1&1&0&0&1\\
\hline
0&0&0&1&0&1&1&0\\
\hline
\end{array}
&\rightarrow &
\begin{array}{|c|c|c|c|c||c|c|c|}
\hline
1&0&0&\rc{0}&1&0&1&0\\
\hline
\rc{0}&1&0&0&1&0&1&\rc{0}\\
\hline
0&0&\rc{0}&1&1&0&0&1\\
\hline
1&1&0&0&\rc{1}&1&1&1\\
\hline\hline
0&\rc{0}&0&0&0&0&0&0\\
\hline
\end{array}
\end{eqnarray*}

We can see that in the array in the right, every line of slope 0 and
1 with its corresponding
independent parity bit has even parity, while every line of slope 2
with its corresponding independent parity bit has odd parity
(this case illustrated in red for the line through entry $c_{1,0}$).
}
\end{example}

\begin{cor}
\label{cor00}
{\em If code $IP(p,r,q)$ is MDS, then
code $EIP(p,r,q,g(x))$ 
given by
Definition~\ref{defEEO} is also MDS.
}
\end{cor}
\pf Similar to Corollary~\ref{cor0}. \qed

Contrary to $BR$ codes, $IP$ codes are not always MDS.
In particular, $IP(p,r,q)$ codes, and hence, by
Corollary~\ref{cor00},\\ $EIP(p,r,q,g(x))$ codes, are MDS for $1\leq
r\leq 3$~\cite{bbbm,bbv}.
For $r>3$ the codes are MDS depending on the prime number chosen.
A list of prime numbers for which $IP(p,r,q)$ is MDS and $r\geq 4$ is
given in~\cite{bbv}. See also~\cite{hsl}.

In the definitions of codes $EBR(p,r,q,g(x))$ and $EIP(p,r,q,g(x))$,
it is assumed that 
each column in an array in the code is in the cyclic code
$\C(p,g(x)(1\xor x),q,d)$. It is 
certainly possible to extend the definition such that each data
column $j$ is in a cyclic code $\C(p,g_j(x)(1\xor x),q,d_j)$, where
each $g_j(x)$ divides $1\xor x^p$ and $\gcd(g_j(x),1\xor x)\eq 1$. In
this case, up to $d_j-1$ erasures can be corrected in data column $j$, giving
unequal erasure correction for the data columns. The $r$ parity columns are
in $\C(p,g'(x)(1\xor x),q,d')$, where $g'(x)$ is the greatest common
divisor of the $g_j(x)$s.

Before proceeding further, we briefly discuss the applications and
advantages of expanded array codes over other array codes.
The applications of expanded array codes are the same as the ones of
traditional array codes,
like BR~\cite{br}, EVENODD~\cite{bbbm}, IP~\cite{bbbmv,bbv},
RDP~\cite{Cor+04}, generalized RDP~\cite{b,f} and codes with
distributed parities~\cite{p,xbbw,xb,zzs}. Mainly, these array codes
can be used in RAID type of
architectures~\cite{g}, like RAID~6, which requires two parity
columns. In addition, expanded array codes contain vertical parities,
providing for local recovery~\cite{ghsy}. Array codes are an
alternative to RS codes~\cite{ms}, which
require finite field operations. Since array codes are based on XOR
operations, in general their implementation has less complexity than the
one of codes based on finite fields.

Another application of array codes is the cloud. In this case,
each entry may correspond to a whole device. Erasure codes involving
local and global parities may be invoked~\cite{ghjy,ghsy,ghswy}. Array
codes naturally provide horizontal locality as well as column
recovery, so they can be used as Locally Recoverable (LRC) Codes.
Traditional array codes do not have vertical parity. In some
applications, a column may represents a device, and each entry in the
column a sector or a page in the device. It may be desirable to have
vertical parities in an array code, so, if a number of sectors or
pages fail (for instance, if
their internal ECCs are exceeded and such a situation is detected by
the CRCs), the failed sectors or pages can be recovered locally without
invoking other devices (a first responder type of approach). A way to
achieve this goal is by using an array code like a BR or an
IP code and then encoding each column with
vertical parities. A problem with this approach is that
the column parities are not protected by the other parities. This
problem is overcome by the expanded array codes described above.

$EBR(p,r,q,g(x))$ codes have another interesting property. In
addition to being able to recover any $r$ erased columns, they can
recover also from a number of erased lines of
slope $j$, $1\leq j\leq r-1$. We will study this property in some
detail in Section~\ref{lines}, but in the meantime we illustrate this
property with a simple example.

\begin{example}
\label{ex0}
{\em
Assume that $p\eq 5$, and we have a $5\times 5$ array such
that we encode a $4\times 3$ data array consisting of zeros into a
$4\times 5$ array in code $BR(5,2,2)$, and then we append a parity row. The result is a
$5\times 5$ 0-array. Next, assume that two lines of slope 1 are erased, say,
lines 1 (in blue) and 4 (in red) as follows, where the symbol $E$ corresponds to an
erasure:

$$
\begin{array}{|c|c|c|c|c|}
\hline
0&\bc{E}  & 0& 0&\rc{E}  \\\hline
\bc{E} & 0& 0&\rc{E}  & 0\\\hline
0& 0&\rc{E}  & 0&\bc{E}  \\\hline
0&\rc{E}  & 0&\bc{E}  & 0\\\hline
\rc{E} & 0&\bc{E}  & 0& 0\\
\hline
\end{array}
$$

This situation admits two solutions as shown below, since the top 4
rows in the array in the right are in $B(5,2,2)$:

$$
\begin{array}{cc}
\begin{array}{|c|c|c|c|c|}
\hline
0& \bc{0}& 0& 0& \rc{0}\\\hline
\bc{0}& 0& 0& \rc{0}& 0\\\hline
0& 0& \rc{0}& 0& \bc{0}\\\hline
0& \rc{0}& 0& \bc{0}& 0\\\hline
\rc{0}& 0& \bc{0}& 0& 0\\
\hline
\end{array}
&
\begin{array}{|c|c|c|c|c|}
\hline
0& \bc{1} &0 & 0& \rc{1} \\\hline
\bc{1} & 0& 0& \rc{1} & 0\\\hline
0& 0& \rc{1} & 0& \bc{1} \\\hline
0& \rc{1} & 0& \bc{1} & 0\\\hline
\rc{1}& 0& \bc{1} & 0& 0\\
\hline
\end{array}
\end{array}
$$

Hence, the two erased lines of slope 1 are uncorrectable.
If we encode the $4\times 3$ data array of zeros into a $5\times 5$
array in $EBR(5,2,2,1)$, we also obtain the zero array, but there is
a unique decoding to the two erased lines of slope 1, given by the
zero array in the left. Since the two colored diagonals
in the array in the right have odd parity, this array cannot be
in $EBR(5,2,2,1)$ and the solution is unique.
}
\end{example}


If each entry represents a page or a
sector, for example, a 64K sector, the parity sectors of an expanded
array code, being obtained
as XORs of data sectors, by linearity, inherit the CRC bits, i.e.,
the CRC of the parity sectors does not need to be computed. This one
is an important advantage in implementation. Finally, expanded array codes
naturally provide multiple localities to recover a single failed
symbol, a problem that attracted attention in recent
literature~\cite{tb,tb2,zy}.

The paper is structured as follows: in Section~\ref{EBR}, we present
efficient encoding and decoding algorithms for the array
codes we have defined above (i.e., EBR and EIP codes). In
Section~\ref{distance}, we examine the problem of the minimum
(symbol, as opposed to column) distance of such array codes. In
Section~\ref{lines}, we study 
conditions under which erased lines of slope $s$,
$0\leq s\leq r-1$, can be recovered in EBR codes (as illustrated in
Example~\ref{ex0} for $s\eq 1$).
Section~\ref{punctured} discusses the puncturing of EBR and EIP codes
to obtain MDS codes consisting of
$m\times p$ or $m\times (p+r)$
arrays, where $m<p$ for certain values of
$m$. We end the paper by drawing some conclusions.

\section{Encoding, Decoding and Updating of a Data Symbol in EBR and EIP Codes}
\label{EBR}
We start with a technical lemma.

\begin{lemma}
\label{l8}
{\em
Let $g(x)$
be an irreducible polynomial on $GF(q)$ such that $g(x)$ divides
$x^p\xor 1$, $p$ prime, and
$\gcd(g(x),x\xor 1)\eq 1$. Then, for each $i$ such that $1\leq i<p$,
$\gcd(g(x),x^i\xor 1)\eq 1$.
}
\end{lemma}
\pf Assume that the lemma is not true, hence, since $g(x)$ is
irreducible, there is an $i$, $1\leq i<p$, such that $g(x)$ divides
$x^i+1$. Moreover, assume that $i$ is minimal with this property.
Since $\gcd(g(x),x\xor 1)\eq 1$,
then $2\leq i<p$. Let $p\eq ci+r$, where, since $p$ is
prime, $1\leq r<i$. We can easily verify that

\begin{eqnarray}
\label{eq15}
x^p\xor 1&=&(x^i\xor 1)\left(\bigoplus_{j=1}^cx^{p-ji}\right)\xor x^r\xor 1.
\end{eqnarray}

Since $g(x)$ divides both $x^p\xor 1$ and $x^i\xor 1$, then, by~(\ref{eq15}), it also
divides $x^r\xor 1$, contradicting the minimality of $i$.
\qed

The following lemma gives a recursion
that will be used in the decoding of EBR codes. 

\begin{lemma}
\label{l1}
{\em
Let $\uv(\al)\eq\bigoplus_{i=0}^{p-1}v_i\al^i\in\C(p,g(x)(1\xor
x),q,d)$, where 
$\al\in\R_p(q)$, $\al\neq 1$ and $\al^p\eq 1$.
Then, for each $j$ such that $1\leq j\leq p-1$, the recursion
$(1\xor\al^j)\uz(\al)\eq \uv(\al)$ has a unique
solution in $\C(p,g(x)(1\xor x),q,d)$.
Specifically, if
$\uz(\al)\eq\bigoplus_{i=0}^{p-1}z_i\al^i$,
then

\begin{eqnarray}
\label{eq0}
z_0&=&\bigoplus_{u=1}^{(p-1)/2}v_{\lan 2uj\ran}\\
\label{eq1}
z_{\lan ij\ran}&=&z_{\lan (i-1)j\ran}\xor v_{\lan ij\ran}\quad {\rm for}\quad 1\leq i\leq p-1.
\end{eqnarray}
\qed
}
\end{lemma}

\noindent\pf
If $z_0$ is known, by solving recursively $(1\xor\al^j)\uz(\al)\eq \uv(\al)$,
(\ref{eq1}) is obtained.
Since $p$ is prime, all the entries $z_i$
for $1\leq i\leq p-1$ are covered by this recursion. In particular,
if $i\eq 1$, $z_j\eq z_0\xor v_j$. Using this result as our starting
point, we obtain  

\begin{eqnarray}
\label{eq7}
z_{\lan ij\ran}&=&z_0\xor\bigoplus_{u=1}^{i}v_{\lan uj\ran}\quad {\rm
for}\quad 1\leq i\leq p-1.
\end{eqnarray}

XORing both sides of~(\ref{eq7}) from $i\eq 1$ to $i\eq p-1$, we have

\begin{eqnarray}
\label{eq9}
\bigoplus_{i=1}^{p-1}z_{\lan
ij\ran}\quad\eq\quad\bigoplus_{i=1}^{p-1}z_i&=&(p-1)z_0\xor\bigoplus_{i=1}^{p-1}\bigoplus_{u=1}^{i}v_{\lan
uj\ran}
\end{eqnarray}

Since, in particular,  
$\uz(\al)$ must have even weight, then
$\bigoplus_{i=1}^{p-1}z_i\eq z_0$. Also, since $p$ is odd,
$(p-1)z_0\eq 0$. Finally,

\begin{eqnarray*}
\bigoplus_{i=1}^{p-1}\bigoplus_{u=1}^{i}v_{\lan uj\ran}&\eq &
(p-1)v_j\xor (p-2)v_{\lan 2j\ran}\xor (p-3)v_{\lan
3j\ran}\xor\cdots\xor 2v_{\lan (p-2)j\ran}\xor v_{\lan
(p-1)j\ran}\\
&=& \bigoplus_{u=1}^{(p-1)/2}v_{\lan 2uj\ran}.
\end{eqnarray*}

Replacing these values in~(\ref{eq9}), we obtain~(\ref{eq0}).

It remains to be proven that $\uz(\al)\in\C(p,g(x)(1\xor x),q,d)$.
Hence, we have to prove that $g(x)(1\xor x)$ divides $\uz(x)$.
Certainly, $1\xor x$ divides $\uz(x)$ since $\uz(x)$ has even weight.
Without loss of generality, assume that $g(x)$ is irreducible
(otherwise, take an irreducible factor of $g(x)$). Since $g(x)$
divides $\uv(x)$, $(1\xor\al^j)\uz(\al)\eq \uv(\al)$
and, by Lemma~\ref{l8}, $\gcd(g(x),1\xor
x^j)\eq 1$ for $1\leq j\leq p-1$, $g(x)$ divides $\uz(x)$.
\qed

Lemma~\ref{l1} was proven in~\cite{hhsl}, Lemma~7, and in~\cite{hscl},
Lemma~13, for the special case $\C(p,1\xor x,2,2)$.

The next example illustrates Lemma~\ref{l1}:

\begin{example}
\label{ex1}
{\em
Let $p\eq 7$, $\uv(\al)\eq
1\xor\al\xor\al^4\xor\al^6\in\C(7,(1\xor x\xor x^3)(1\xor x),2,4)$ (see
Example~\ref{ex8}), i.e.,
$v_0\eq 1$, $v_1\eq 1$, $v_2\eq 0$, $v_3\eq 0$, $v_4\eq 1$, $v_5\eq
0$ and $v_6\eq 1$. Assume that we want to solve the recursion
$(1\xor\al^3)\uz(\al)\eq \uv(\al)$. According to~(\ref{eq0})
and~(\ref{eq1}), since $j\eq 3$, $\lan 2j\ran_7\eq 6$, so

\begin{eqnarray*}
z_0&=&v_6\xor v_{\lan 12\ran_7}\xor v_{\lan 18\ran_7}\eq v_4\xor v_5\xor v_6\;\eq\; 0\\
z_3&=&z_0\xor v_3\;\eq\; 0\\
z_6&=&z_3\xor v_6\;\eq\; 1\\
z_2&=&z_6\xor v_2\;\eq\; 1\\
z_5&=&z_2\xor v_5\;\eq\; 1\\
z_1&=&z_5\xor v_1\;\eq\; 0\\
z_4&=&z_1\xor v_4\;\eq\; 1,\\
\end{eqnarray*}
so $\uz(\al)\eq \al^2\xor\al^4\xor\al^5\xor\al^6$. In particular, we
can see that $\uz\in\C(7,(1\xor x\xor x^3)(1\xor x),2,4)$.
}
\end{example}

Observe that the recursion in Lemma~\ref{l1} involves $\frac{3p-5}{2}$
XORs.

Next we will show how to correct up to $r$ erased columns in $EBR(p,r,q,g(x))$ by
adapting the method in~\cite{br}. The next theorem was proven
in~\cite{bdh} for $EBR(p,r,q,1)$. The proof is analogous, but we give
it for the sake of completeness.

\begin{theorem}
\label{theo1}
{\em
Code $EBR(p,r,q,g(x))$ given by Definition~\ref{defGEBR} can correct up to
$d-1$ erasures or a burst of up to $1+\deg(g(x))$ (consecutive) erasures in each
column and up to $r$ erased columns.
}
\end{theorem}

\noindent\pf Given an array in $EBR(p,r,2,g(x))$, since columns are
in the cyclic code $\C(p,g(x)(1\xor x),q,d)$,
up to $d-1$ erasures can be corrected in each column, and also a
burst of up to the degree of the generator polynomial $g(x)(1\xor
x)$, i.e.,\\ $1+\deg(g(x))$ (systematic encoding is a special
case of recovering such a burst of erasures).

Next assume that columns $i_0,i_1,\ldots,i_{\rho-1}$ have been erased,
where $\rho\leq r$, and we denote by $\ue_s$ the (erased) value of
column $i_s$.
Consider the polynomial $G(x)$ of degree $\rho -1$

\begin{eqnarray}
\label{eq2}
G(x)&=&\prod_{s=1}^{\rho-1}(x\xor\al^{i_s})\eq \bigoplus_{s=0}^{\rho-1}g_sx^s.
\end{eqnarray}

Notice that
\begin{eqnarray}
\label{eq3}
G(\al^{i_0})&=&\prod_{s=1}^{\rho-1}(\al^{i_0}\xor\al^{i_s})\\
\label{eq4}
G(\al^{i_j})&=&0\quad {\rm for}\quad 1\leq j\leq\rho -1. 
\end{eqnarray}

Denote the columns of the array by $\uc_u$, where $0\leq u\leq p-1$.
Assuming that the erased columns are zero, compute the syndromes

\begin{eqnarray}
\label{eq5}
S_j&=&\bigoplus_{u=0}^{p-1}\al^{ju}\uc_u\quad {\rm for}\quad 0\leq j\leq\rho-1.
\end{eqnarray}

Hence, from~(\ref{eq5}), we also have

\begin{eqnarray}
\label{eq6}
S_j&=&\bigoplus_{s=0}^{\rho-1}\al^{ji_s}\ue_{s}\quad {\rm for}\quad 0\leq j\leq\rho-1.
\end{eqnarray}



From~(\ref{eq2}), (\ref{eq3}), (\ref{eq4}) and (\ref{eq6}), 

\begin{eqnarray}
\nonumber
\bigoplus_{j=0}^{\rho-1}g_{j}S_{j}&=&\bigoplus_{j=0}^{\rho-1}g_{j}\bigoplus_{s=0}^{\rho-1}\al^{ji_s}\ue_{s}\\
\nonumber
&=&\bigoplus_{s=0}^{\rho-1}\ue_{s}\bigoplus_{j=0}^{\rho-1}g_{j}(\al^{i_s})^j\\
\nonumber
&=&\bigoplus_{s=0}^{\rho-1}\ue_{s}G(\al^{i_s})\\
\label{eq8}
&=&\left(\prod_{s=1}^{\rho-1}(\al^{i_0}\xor\al^{i_s})\right)\ue_0.
\end{eqnarray}

After computing $\bigoplus_{j=0}^{\rho-1}g_{j}S_{j}$, $\ue_0$ can
be obtained by applying the recursion given by~(\ref{eq0})
and~(\ref{eq1}) in Lemma~\ref{l1} $\rho -1$ times. Once $\ue_0$ is
obtained, we are left with $\rho -1$ erasures, and we proceed by
induction.
\qed

The next example illustrates the decoding procedure given in
Theorem~\ref{theo1}.

\begin{example}
\label{ex2}
{\em
Consider the code $EBR(7,3,2,1\xor x\xor x^3)$ of Example~\ref{ex8} and
assume that we want to decode the following
array, where the blank spaces denote erasures::

$$
\begin{array}{|c|c|c|c||c|c|c|}
\hline
&\phantom{1}&&\phantom{1}&1&0&\phantom{1}\\
\hline
1&&&&&0&\\
\hline
&&1&&0&&\\
\hline\hline
0&&0&&&&\\
\hline
1&&0&&0&&\\
\hline
&&&&1&&\\
\hline
1&&&&&1&\\
\hline
\end{array}
$$

We can see that columns 1, 3 and 6 are erased, columns 0 and 4
contain three erasures each and columns 2 and 5 contain a burst of
length four (in particular, the burst in column~2 is an all-around
burst, but it can be corrected also since the code is cyclic). Since
the columns are in the cyclic
code $\C(7,(1\xor x\xor x^3)(1\xor x),2,4)$, the first step is
obtaining the erasures in columns 0, 2, 4 and~5.
Once this is done, we obtain

$$
\begin{array}{|c|c|c|c||c|c|c|}
\hline
1&\phantom{1}&1&\phantom{1}&1&0&\phantom{1}\\
\hline
1&&1&&0&0&\\
\hline
0&&1&&0&1&\\
\hline\hline
0&&0&&1&0&\\
\hline
1&&0&&0&1&\\
\hline
0&&1&&1&1&\\
\hline
1&&0&&1&1&\\
\hline
\end{array}
$$

By (\ref{eq2}),

\begin{eqnarray}
\label{eq50}
G(x)&=&g_0\xor g_1x\xor g_2x^2\;\eq\;(x\xor\al^3)(x\xor\al^6)\;\eq\;\al^2\xor (\al^3\xor\al^6)x\xor x^2.
\end{eqnarray}

Assuming that the erased columns are zero when computing the
syndromes, by~(\ref{eq5}), we obtain

\begin{eqnarray*}
S_0&=&1\xor\al^3\xor\al^5\xor\al^6\\
S_1&=&\al\xor\al^2\xor\al^3\xor\al^6\\
S_2&=&1\xor\al\xor\al^4\xor\al^6.\\
\end{eqnarray*}

Using~(\ref{eq50}), we compute

\begin{eqnarray}
\label{eq51}
g_0S_0\xor g_1S_1\xor g_2S_2 &=&1\xor\al\xor\al^2\xor\al^5.
\end{eqnarray}

By~(\ref{eq8}) and~(\ref{eq51}), we have to solve the double recursion

\begin{eqnarray*}
(\al\xor\al^3)(\al\xor\al^6)\ue_0 &=&1\xor\al\xor\al^2\xor\al^5,
\end{eqnarray*}
or, multiplying both sides by $\al^{-2}$,

\begin{eqnarray*}
(1\xor\al^2)(1\xor\al^5)\ue_0 &=&1\xor\al^3\xor\al^5\xor\al^6.
\end{eqnarray*}

Let $(1\xor\al^5)\ue_0\eq\uv_0$, then we have to solve first

\begin{eqnarray*}
(1\xor\al^2)\uv_0 &=&1\xor\al^3\xor\al^5\xor\al^6.
\end{eqnarray*}

Applying the recursion given by~(\ref{eq0}) and~(\ref{eq1}) as
illustrated in Example~\ref{ex2}, we obtain,

\begin{eqnarray*}
\uv_0 &=&1\xor\al^2\xor\al^3\xor\al^4.
\end{eqnarray*}

Next we have to solve

\begin{eqnarray*}
(1\xor\al^5)\ue_0 &=&1\xor\al^2\xor\al^3\xor\al^4.
\end{eqnarray*}

This gives,

\begin{eqnarray*}
\ue_0 &=&\al\xor\al^2\xor\al^3\xor\al^6.
\end{eqnarray*}

Recomputing the syndromes,

\begin{eqnarray*}
S_0 &=&S_0\xor \ue_0\;\eq\;1\xor\al\xor\al^2\xor\al^5\\
S_1 &=&S_1\xor \al \ue_0\;\eq\;1\xor\al\xor\al^4\xor\al^6.\\
\end{eqnarray*}

Repeating the procedure for two erasures, we now have

\begin{eqnarray*}
G(x)&=&g_0\xor g_1x\;\eq\;\al^6\xor x,
\end{eqnarray*}
which gives

\begin{eqnarray*}
g_0S_0\xor g_1S_1 &=&0.
\end{eqnarray*}

We have to solve the recursion

\begin{eqnarray*}
(\al^3\xor\al^6)\ue_1 &=&0,
\end{eqnarray*}
hence,

\begin{eqnarray*}
\ue_1 &=&0.
\end{eqnarray*}

Finally, we recompute

\begin{eqnarray*}
S_0 &=&S_0\xor \ue_1\;\eq\;1\xor\al\xor\al^2\xor\al^5\;\eq\;\ue_2.\\
\end{eqnarray*}

The final decoded array is then

$$
\begin{array}{|c|c|c|c||c|c|c|}
\hline
1&0&1&0&1&0&1\\
\hline
1&1&1&0&0&0&1\\
\hline
0&1&1&0&0&1&1\\
\hline\hline
0&1&0&0&1&0&0\\
\hline
1&0&0&0&0&1&0\\
\hline
0&0&1&0&1&1&1\\
\hline
1&1&0&0&1&1&0\\
\hline
\end{array}\;\;,
$$
which coincides with the first array given in Example~\ref{ex8}.
}
\end{example}

The encoding is a special case of the decoding. For
example, we may use the last $1+\deg(g(x))$ rows and the last $r$ columns to store
the parities. We start by encoding systematically~\cite{ms} the first
$p-r$ columns using the generator polynomial $g(x)(1\xor x)$. Next we compute
the last $r$ columns using the decoding procedure as described in
Theorem~\ref{theo1}. Since at the encoding the erasures are always in
the last $r$ columns, we may precompute the coefficients of
$G(x)\eq\prod_{j=1}^{r-1}(x\xor\al^{p-r+j})$, making the process
faster.

Let us examine next $EIP(p,r,q,g(x))$ codes. The encoding of
$EIP(p,r,q,g(x))$ codes is very simple and is a direct consequence of
Definition~\ref{defEEO}: given a $(p-(1+\deg(g(x)))\times p$ data array that we
denote as $\uv_{0},\uv_{1},\ldots,\uv_{p-1}$, each $\uv_j$ a
(vertical) vector of length $p-(1+\deg(g(x))$ over $GF(q)$, we start
by encoding systematically each $\uv_j$ into $\uc_j\in\C(p,g(x)(1\xor x),q,d)$. The
result is a $p\times p$ array. Then we obtain the $r$ parity columns
$\uc_{p+s}$, $0\leq s\leq r-1$, as 
%
$\uc_{p+s}\eq\bigoplus_{j=0}^{p-1}\al^{sj}\uc_j$.
Thus, once
the $\uc_j$s have been obtained, each $\uc_{p+s}$ requires $(p-1)p$
XORs.

Let us consider next the special case of
$EIP(p,r,q,1)$ codes. In a first step, we need to obtain symbols
$c_{p-1,j}$ for\\ $0\leq j\leq p-1$ as the XOR of symbols $c_{i,j}$ for
$0\leq i\leq p-2$, which takes $p-2$ XORs for each $c_{p-1,j}$.
Hence, the total number of XORs required by the encoding algorithm is
$(p-2)p+r(p-1)p$. If
we shorten the code to $k$ columns, where $1\leq k\leq p$, by
assuming that $p-k$ of the columns are zero, then the total number of
XORs required at the encoding is $k(p-2)+r(k-1)p$. In particular, if $r\eq 2$
and $k\leq p-2$, the total number of XORs at the encoding is $3kp-2(k+p)$.
The number of XORs according to the optimized encoding
algorithm for $EBR(p,2,q,1)$ with $k$ data columns given in~\cite{bdh}  is
$3kp-(k+2)$, so the encoding algorithm for $EIP(p,2,q,1)$ also with
$k$ data columns is more efficient.

Table~\ref{t1} compares the number of XORs of required by the
encoding algorithms
of $EIP(p,2,q,1)$ and of $EBR(p,2,q,1)$ as given in~\cite{bdh}, both
codes shortened to $k$ data columns for $1\leq k\leq p-2$.
The encoding algorithm of $EIP(p,r,q,1)$
always requires less XORs than the optimized encoding
algorithm of $EBR(p,2,q,1)$ in~\cite{bdh}. Table~\ref{t1} shows
that the savings are more dramatic when $k<<p$.

\begin{table}
\begin{center}
\begin{tabular}{|c|c|c|c|c|}
\hline
&&Encoding Algorithm from~\cite{bdh}&Encoding Algorithm for $EIP(p,2,q,1)$&\\
\hline
$p$&$k$ &$3kp-(k+2)$&$3kp-2(k+p)$&Improvement \%\\
\hline\hline
17&8&398&358&10.1\%\\
\hline
17&15&748&701&6.3\%\\
\hline
127&8&3038&2778&8.6\%\\
\hline
127&50&18998&18696&1.6\%\\
\hline
127&125&47498&47121&.8\%\\
\hline
257&8&6158&5638&8.4\%\\
\hline
257&50&38498&37936&1.5\%\\
\hline
257&255&196348&195581&.4\%\\
\hline
\end{tabular}
\end{center}
\caption{\label{t1} Comparison between the number of XORs of
optimized encoding
algorithm for $EBR(p,2,q,1)$ in~\cite{bdh} and Encoding Algorithm for
$EIP(p,2,q,1)$ with $k$ data columns, $1\leq k\leq p-2$.}
\end{table}

Regarding the decoding of $EIP(p,r,q,g(x))$ codes, the first step is
always correcting up to $d-1$ erasures in each column or a burst of
up to $1+\deg(g(x))$ erasures wherever this is
feasible. Once this is done, assuming that the code is MDS and up to
$r$ columns are erased, we can apply transform~(\ref{eq100}) and decode
the array in $IP(p,r,q)$. Then the inverse transformation will give the
desired array in $EIP(p,r,q,g(x))$.

If the erased columns correspond
to data columns, i.e., they are among the first $p$ columns in the array,
the array can be decoded directly in $EIP(p,r,q,g(x))$ applying the
same method as for the decoding of $EBR(p,r,q,g(x))$. If some of the
erased columns are among the $r$ parity columns, the decoding is more
complicated since the recursion of Lemma~\ref{l1} cannot be applied.
The case $r\eq 2$ is simple to handle though, since when one of the
two parity columns is erased, it is corrected as a special case.

We end this section with the problem of updating $EBR(p,r,q,g(x))$
and $EIP(p,r,q,g(x))$ codes. The idea is, when updating one data
symbol, how to minimize the number of parity symbols that need to be
updated, a problem that has been treated repeatedly in the literature
on array codes~\cite{bbbm,bbv,br2,p,xbbw,xb}. Actually,
$EBR(p,r,q,g(x))$ codes have bad updating properties, since the
parities are not independent and updating one data symbol causes the
updating of most of the parity symbols. The same is true for
$BR(p,r,q)$ codes, and the creation of codes like EVENODD~\cite{bbbm}
arises from the need of optimizing the number of updates by making
the parities independent. Hence, in what follows, we concentrate on
$EIP(p,r,q,g(x))$ codes only.

As usual, denote an array in $EIP(p,r,q,g(x))$ as
$(\uc_0,\uc_1,\ldots,\uc_{p-1},\uc_p,\uc_{p+1},\ldots,\uc_{p+r-1})$,
where each $\uc_j$ is a (column) vector of length $p$.
Each time a data symbol $c_{i,j}$, 
$0\leq i\leq p-(2+\deg(g(x)))$, $0\leq j\leq p-1$, is updated,
first we need to update the parity symbols in column $j$. In effect,
if data symbol $c_{i,j}$
is replaced by symbol $c'_{i,j}$, consider the (vertical) vector $\uv_j$ of
length $p-(1+\deg(g(x)))$ that is zero everywhere except in location
$i$, where it is equal to $c_{i,j}\xor c'_{i,j}$. Encoding
(systematically) $\uv_j$ into $\C(p,g(x)(1\xor x),q,d)$, we
obtain a (vertical) vector 
that we denote $\uc'_j$.
Once $\uc'_j$ is obtained, we update $\uc_j$ as $\uc_j\xor\uc'_j$ and
the parity vectors $\uc_{p+s}$, $0\leq s\leq r-1$, as
$\uc_{p+s}\eq\uc_{p+s}\xor\al^{sj}\uc'_j$. Let us illustrate the
process in the next example.

\begin{example}
\label{ex17}
{\em
Consider the following array in code $EIP(7,3,2,1\xor x\xor x^3)$:

$$
\begin{array}{|c|c|c|c|c|c|c||c|c|c|}
\hline
1&0&0&1&0&0&1&1&0&0\\
\hline
1&1&0&0&1&0&1&0&0&0\\
\hline
1&1&1&0&1&1&1&0&0&1\\
\hline\hline
0&1&0&1&1&0&0&1&0&0\\
\hline
0&0&1&0&0&1&0&0&0&1\\
\hline
1&0&1&1&0&1&1&1&0&1\\
\hline
0&1&1&1&1&1&0&1&0&1\\
\hline
\end{array}
$$

Assume that we want to update symbol $c_{2,1}$. The first step is
encoding (systematically) $\uv_1\eq (0,0,1)$ in\\
$\C(7,(1\xor x\xor x^3)(1\xor x),2,4)$.
Doing so, we obtain $\uc_1'\eq (0,0,1,0,1,1,1)$ and
$\uc_1\xor\uc'_1\eq (0,1,0,1,1,1,0)$.

Then, $\uc_7\xor\uc'_1\eq (1,0,1,1,1,0,0)$,
$\uc_8\xor\al\uc'_1\eq (1,0,0,1,0,1,1)$ and
$\uc_9\xor\al^2\uc'_1\eq (1,1,1,0,0,1,0)$. The updated array is then

$$
\begin{array}{|c|c|c|c|c|c|c||c|c|c|}
\hline
1&0&0&1&0&0&1&1&1&1\\
\hline
1&1&0&0&1&0&1&0&0&1\\
\hline
1&0&1&0&1&1&1&1&0&1\\
\hline\hline
0&1&0&1&1&0&0&1&1&0\\
\hline
0&1&1&0&0&1&0&1&0&0\\
\hline
1&1&1&1&0&1&1&0&1&1\\
\hline
0&0&1&1&1&1&0&0&1&0\\
\hline
\end{array}\;\;.
$$
}
\end{example}

We can see that the lowest number of updates in the parity symbols
that an $EIP(p,r,q,g(x))$ code as given by Definition~\ref{defIP} 
can make is $(r+1)d-1$. In Example~\ref{ex17},
this is the case, we are updating $(3)(4)-1\eq 11$ parity symbols.
The reason is that the three vectors consisting of the systematic
encoding of the three vectors of weight 1 and length 3 in the
vertical code $\C(7,(1\xor x\xor x^3)(1\xor x),2,4)$, when encoded
systematically, have weight 4, the minimum distance of the code. Let us state this
observation as a lemma.

\begin{lemma}
\label{l9}
{\em
Consider an $EIP(p,r,q,g(x))$ with vertical code $\C(p,g(x)(1\xor
x),q,d)$. Then the number of parity updates when a data symbol is updated
reaches the optimal value $(r+1)d-1$ if and only if the systematic
encoding of each vector of weight one and length $p-(1+\deg(g(x))$
has weight $d$.
\qed
}
\end{lemma}

\begin{cor}
\label{cor4}
{\em
Consider an $EIP(p,r,q,g(x))$ code and assume that the vertical code
$\C(p,g(x)(1\xor x),q,d)$ is MDS. Then the number of parity updates
when a data symbol is updated reaches the optimal value $(r+1)d-1$.
}
\end{cor}
\pf Simply observe that if $\C(p,g(x)(1\xor x),q,d)$ is MDS then the
systematic encoding of each vector of weight 1 and length $p-(1+\deg(g(x))$ has
weight $2+\deg(g(x))\eq d$ and the result follows from Lemma~\ref{l9}.
\qed

\begin{cor}
\label{cor4bis}
{\em
Consider an $EIP(p,r,2,1)$ code. Then the number of parity updates
when a data symbol is updated reaches the optimal value $2r+1$.
}
\end{cor}
\pf This is the special case of Corollary~\ref{cor4} corresponding to
the binary field $GF(2)$ and the vertical code is the\\
$\C(p,1\xor x,2,2)$ parity code, which in particular is MDS.
\qed

\section{Minimum Hamming Distance of Array Codes with Local Properties}
\label{distance}
In this section we consider a problem that has not received much
attention in the literature on array codes. In general, when we talk
about the distance of an array code, we refer to the column distance.
In this section we want to consider the symbol distance. Having a
high symbol distance may be important when erased columns co-exist
with erased symbols. We will consider both $EBR(p,r,q,g(x))$ and
$EIP(p,r,q,g(x))$ codes. We have already seen that $EBR(p,r,q,g(x))$
codes are MDS, i.e., their column distance is $r+1$, while
$EIP(p,r,q,g(x))$ are MDS depending on the prime number considered
and the value of $r$. We will simply call the
Hamming distance the symbol distance of a code, otherwise we refer to
the column distance.

Let us start with a lower bound. 

\begin{lemma}
\label{l2}
{\em
Let $D$ be the Hamming distance of an $EBR(p,r,q,g(x))$ or an
$EIP(p,r,q,g(x))$ MDS code.
Then, $D\geq d(r+1)$.
}
\end{lemma}

\noindent\pf
Take a non-zero array in $EBR(p,r,q,g(x))$ or in $EIP(p,r,q,g(x))$.
Then, since the code is MDS, at least $r+1$ columns in
the array are non-zero. Since each non-zero column has weight at
least $d$, the result follows. \qed

Finding $D$ for an $EIP(p,r,q,g(x))$ MDS code is easy, as shown in
the next corollary.

\begin{corollary}
\label{cor5}
{\em
Consider an $EIP(p,r,q,g(x))$ MDS code with minimum Hamming distance
$D$. Then, $D\eq d(r+1)$.
}
\end{corollary}
\pf Take a $p\times p$ array consisting of a column of weight
$d$ in $\C(p,g(x)(1\xor x),q,d)$, while the remaining $p-1$ columns are
zero. Encoding this array in $EIP(p,r,q,g(x))$, since the parity
columns are rotations of the non-zero data column, they also have weight $d$,
so we obtain an array
with $r+1$ columns of weight $d$, hence $D\leq d(r+1)$. The result then
follows from Lemma~\ref{l2}.
\qed

From now on, we consider $EBR(p,r,q,1)$ codes only.

\begin{lemma}
\label{cor1}
{\em
Consider the code $EBR(p,r,q,1)$, where either $1\leq
r\leq 3$ or $p-2\leq r\leq p-1$. Then the minimum Hamming
distance of $EBR(p,r,q,1)$ is $D\eq 2(r+1)$.
}
\end{lemma}

\noindent\pf
Since $d\eq 2$,
by Lemma~\ref{l2}, $D\geq 2(r+1)$. So, it is enough to exhibit an array of
weight $2(r+1)$ in $EBR(p,r,q,1)$ when $1\leq r\leq 3$ or $p-2\leq r\leq
p-1$. The case $r\eq 1$
is trivial.

Denoting the entries of an array by $(i,j)$, where $0\leq i,j\leq
p-1$, consider the array that is 1 in entries
$(p-3,p-2)$,\\ $(p-3,p-1)$, $(p-2,p-3)$, $(p-2,p-1)$, $(p-1,p-3)$ and $(p-1,p-2)$
and 0 elsewhere. This array is in $EBR(p,2,q,1)$ and it has Hamming
weight 6.

Similarly, consider the array that is 1 in entries
$(p-5,p-2)$, $(p-5,p-1)$, $(p-4,p-3)$, $(p-4,p-1)$, $(p-2,p-4)$, $(p-2,p-2)$, $(p-1,p-4)$ and $(p-1,p-3)$
and 0 elsewhere. This is an array in $EBR(p,3,q,1)$ and it has Hamming
weight 8.

Next take $r\eq p-2$. Consider the array $(c_{i,j})$, $0\leq i,j\leq
p-1$, of Hamming weight $2(p-1)$, such that:

\begin{eqnarray*}
c_{i,i+1}&=&1\quad {\rm for}\quad 0\leq i\leq p-2\\
c_{i,\lan (i+1)/2\ran}&=&1\quad {\rm for}\quad 0\leq i\leq p-2\\
c_{i,j}&=&0\quad {\rm elsewhere}.
\end{eqnarray*}

By construction, since $i+1\neq \lan (i+1)/2\ran$ for $0\leq i\leq
p-2$, each of the
first $p-1$ rows has exactly two 1s, while the last row is zero.
Similarly, the first column is zero while the last $p-1$ columns have
two 1s each, hence, the array has even parity on rows and columns and
weight $2(p-1)$. It remains to be proven that each line of slope $j$,
$1\leq j\leq p-3$, has even parity.

In effect, for each $w$ such that $0\leq w\leq p-1$, take the line
of slope $j$, $0\leq j\leq p-3$, through entry $c_{w,0}$.
By Definition~\ref{def0}, the entries in this
line are $c_{\lan w-jv\ran,v}$ for $0\leq v\leq p-1$. Take first
$w<p-1$, then there is a unique $i$ such that $w-jv\eq i$ and $v\eq
i+1$, mainly, $i\eq \lan (w-j)/(j+1)\ran$ (notice that this is
possible since $j\neq p-1$). Similarly, there is a unique $i$ such
that $w-jv\eq i$ and $v\eq
\lan (i+1)/2\ran$, mainly, $i\eq \lan (2w-j)/(j+2)\ran$ (notice that this is
possible since $j\neq p-2$). Hence, any line of slope $j$, $1\leq
j\leq p-3$, starting at $c_{w,0}$ for $0\leq w\leq p-2$, has exactly
two 1s and thus even parity. If $w\eq p-1$,
the lines of slope $j$ starting at entry $c_{p-1,0}$ can be shown
to contain no 1s, hence they also
have even parity, so the array is in $EBR(p,p-2,q,1)$ and has Hamming
weight $2(p-1)$.

If $r\eq p-1$, consider the following array $(c_{i,j})$, $0\leq i,j\leq
p-1$, of Hamming weight $2p$:

\begin{eqnarray*}
c_{i,i}&=&1\quad {\rm for}\quad 0\leq i\leq p-1\\
c_{i,\lan i+1\ran}&=&1\quad {\rm for}\quad 0\leq i\leq p-1\\
c_{i,j}&=&0\quad {\rm elsewhere}.
\end{eqnarray*}

Using the same methods as above, we can see that each line of slope $j$,
$0\leq j\leq p-2$, contains exactly two 1s, and hence it has even
parity, so the array is in $EBR(p,p-1,q,1)$ and has Hamming
weight $2p$.
\qed

\begin{example}
\label{ex3}
{\em
Consider $p\eq 7$. The following are arrays of weight 4, 6, 8, 10 and
12 in $EBR(7,1,2,1)$, $EBR(7,2,2,1)$, $EBR(7,3,2,1)$, $EBR(7,5,2,1)$
and $EBR(7,6,2,1)$
respectively according to the proof of Lemma~\ref{cor1}:

$$
\begin{array}{ccc}
\begin{array}{|c|c|c|c|c|c|c|}
\hline
0&0&0&0&0&0&0\\\hline
0&0&0&0&0&0&0\\\hline
0&0&0&0&0&0&0\\\hline
0&0&0&0&0&0&0\\\hline
0&0&0&0&0&0&0\\\hline
0&0&0&0&0&1&1\\\hline
0&0&0&0&0&1&1\\\hline
\end{array}
&
\begin{array}{|c|c|c|c|c|c|c|}
\hline
0&0&0&0&0&0&0\\\hline
0&0&0&0&0&0&0\\\hline
0&0&0&0&0&0&0\\\hline
0&0&0&0&0&0&0\\\hline
0&0&0&0&0&1&1\\\hline
0&0&0&0&1&0&1\\\hline
0&0&0&0&1&1&0\\\hline
\end{array}
&
\begin{array}{|c|c|c|c|c|c|c|}
\hline
0&0&0&0&0&0&0\\\hline
0&0&0&0&0&0&0\\\hline
0&0&0&0&0&1&1\\\hline
0&0&0&0&1&0&1\\\hline
0&0&0&0&0&0&0\\\hline
0&0&0&1&0&1&0\\\hline
0&0&0&1&1&0&0\\\hline
\end{array}
\end{array}
$$

$$
\begin{array}{cc}
\begin{array}{|c|c|c|c|c|c|c|}
\hline
0&1&0&0&1&0&0\\\hline
0&1&1&0&0&0&0\\\hline
0&0&0&1&0&1&0\\\hline
0&0&1&0&1&0&0\\\hline
0&0&0&0&0&1&1\\\hline
0&0&0&1&0&0&1\\\hline
0&0&0&0&0&0&0\\\hline
\end{array}
&
\begin{array}{|c|c|c|c|c|c|c|}
\hline
1&1&0&0&0&0&0\\\hline
0&1&1&0&0&0&0\\\hline
0&0&1&1&0&0&0\\\hline
0&0&0&1&1&0&0\\\hline
0&0&0&0&1&1&0\\\hline
0&0&0&0&0&1&1\\\hline
1&0&0&0&0&0&1\\\hline
\end{array}
\end{array}
$$
}
\end{example}

Lemma~\ref{cor1} shows that the bound of Lemma~\ref{l2} is tight
for $EBR(p,r,q,1)$ when $1\leq r\leq 3$ and $p-2\leq r\leq p-1$. What
happens for $4\leq r\leq p-3$? Going back to Example~\ref{ex3}, consider
$EBR(7,4,2,1)$. An exhaustive search shows that the minimum Hamming
distance of $EBR(7,4,2,1)$ is $D\eq 12$, so in this case the bound is not
tight. The following is an array in
$EBR(7,4,2,1)$ of weight 12:

$$
\begin{array}{|c|c|c|c|c|c|c|}
\hline
0&0&0&0&1&0&1\\\hline
0&0&0&0&0&0&0\\\hline
0&0&0&1&1&0&0\\\hline
0&0&0&0&1&1&0\\\hline
0&0&0&0&0&0&0\\\hline
0&0&1&0&1&0&0\\\hline
0&0&1&1&0&1&1\\\hline
\end{array}
$$

Let us point out that a product code of an MDS horizontal code (like
RS) with $r$ parities with a vertical parity code has minimum Hamming
distance $2(r+1)$.

A possible competitor for an $EBR(p,r,q,1)$ code is a code consisting
of $p\times p$ arrays such that their first $p-1$ rows are in $BR(p,r,q)$
and their last row is the XOR of such first
$p-1$ rows. Let us call $BRVP(p,r,q)$ such a code ($BRVP(5,2,2)$ was
illustrated in Example~\ref{ex0}). For example, if
$p\eq 7$, the following is an array in $BRVP(7,3,2)$:

$$
\begin{array}{|c|c|c|c|c|c|c|}
\hline
0&1&1&1&0&1&0\\\hline
0&0&1&0&0&1&0\\\hline
0&0&1&0&1&0&0\\\hline
0&0&0&1&0&1&0\\\hline
1&1&0&0&1&1&0\\\hline
0&1&0&1&0&0&0\\\hline
1&1&1&1&0&0&0\\\hline
\end{array}
$$

We can easily see that both lemmas~\ref{l2} and~\ref{cor1}
apply to $BRVP(p,r,q)$ codes. However, the minimum Hamming distance of
$BRVP(7,4,2)$ is 10, which is less than the minimum Hamming distance 12 of
$EBR(7,4,2,1)$. In effect, notice that the following is an array of
weight 10 in $BRVP(7,4,2)$:

$$
\begin{array}{|c|c|c|c|c|c|c|}
\hline
0&0&0&0&0&1&1\\\hline
0&0&0&0&1&1&0\\\hline
0&0&0&0&0&0&0\\\hline
0&1&0&0&1&0&0\\\hline
0&0&0&0&0&0&0\\\hline
0&0&1&0&0&0&1\\\hline
0&1&1&0&0&0&0\\\hline
\end{array}
$$

Finding the minimum Hamming distance of an $EBR(p,r,q,1)$ code for $4\leq r\leq
p-3$ is an open problem.

\section{Recovery of Erased Lines of slope $i$ in an
$EBR(p,r,q,1)$ Code}
\label{lines}
In Section~\ref{EBR}, we have seen how to encode and decode an
$EBR(p,r,q,g(x))$ code.
In particular, we have shown how to recover up to $r$
erased columns.
Interestingly, an $EBR(p,r,q,1)$ code can also recover a number of erased
lines of slope $i$, where $0\leq i\leq r-1$.
We say that an $EBR(p,r,q,1)$ code is MDS on lines of slope $i$, if the
code can recover up to $r$ erased lines of such slope. We have shown in
Theorem~\ref{theo1} that
an $EBR(p,r,q,1)$ code is MDS on lines of slope $\infty$. What happens
with the other slopes $i$, $0\leq i\leq r-1$?

A trivial case
corresponds to $r\eq 1$: in this case, an $EBR(p,1,q,1)$ code is a
product code with parity on rows and columns. Any erased row or
column can be recovered, hence, an $EBR(p,1,q,1)$ code is MDS on lines of
slope $\infty$ and on lines of slope~0.

The next case corresponds to $r\eq 2$. We had seen in
Example~\ref{ex0} that the code $EBR(5,2,2,1)$ can recover an erased
pair of lines of slope~1. The following lemma proves that this is
true for any $p$.

\begin{lemma}
\label{l3}
{\em
The code $EBR(p,2,q,1)$ can recover any erased pair of
lines of slope $i$ for $0\leq i\leq 1$.
}
\end{lemma}

\noindent\pf
Assume that $(c_{u,v})$, $0\leq u,v\leq p-1$, is an array in
$EBR(p,2,q,1)$, and assume that two rows have been erased. Consider the
transposed array  $(b_{u,v})\eq (c_{u,v})^{\rm T}$, $0\leq u,v\leq
p-1$, i.e., $b_{u,v}\eq c_{v,u}$. Then, the array $(b_{u,v})$ has two
erased columns. It is
enough to show that $(b_{u,v})\in EBR(p,2,q,1)$. Certainly, each line of
slope 0 has even parity. Take a line of slope 1, i.e., according to
Definition~\ref{def0}, $p$ entries
$b_{u,v}$ such that $u+v\eq j$ for some $j$, $0\leq j\leq p-1$. Since
$b_{u,v}\eq c_{v,u}$, the lines of slope 1 coincide for the array
$(c_{u,v})$ and for the transposed array $(b_{u,v})$, so
$(b_{u,v})\in EBR(p,2,q,1)$ and the two erased columns can be recovered.

Assume next that two lines of slope 1 are erased in the array
$(c_{u,v})$.
Take an array $(b_{u,v})$ defined from the array $(c_{u,v})$
as $b_{u,v}\eq c_{\lan -u\ran,\lan u+v\ran}$ for $0\leq u,v\leq
p-1$. We claim, $(b_{u,v})\in EBR(p,2,q,1)$. Notice first that
every line of slope $\infty$ in $(b_{u,v})$ corresponds to a line of
slope 1 in $(c_{u,v})$. In effect, take $v_0$ such that $0\leq
v_0\leq p-1$. According to Definition~\ref{def0}, a line of slope
$\infty$ (vertical) in $(b_{u,v})$ through $b_{0,v_0}$
is given by the set $\{(b_{u,v_0})\,:\,0\leq u\leq p-1\}$, which is
equal to the set $\{(c_{-u,u+v_0})\,:\,0\leq u\leq p-1\}$. This last
set corresponds to the line of slope~1 in $(c_{u,v})$ through
$c_{v_0,0}$, also according to Definition~\ref{def0}.

Similarly, it can be shown that each line of slope 0 in
$(b_{u,v})$ corresponds to a line of slope 0 in $(c_{u,v})$ and
that each line of slope 1 in
$(b_{u,v})$ corresponds to a line of slope $\infty$ in $(c_{u,v})$.
Hence, $(b_{u,v})$ is in $EBR(p,2,q,1)$ since it has even parity on
lines of slope 0, 1 and $\infty$,
so, by Theorem~\ref{theo1}, it can recover the two columns
corresponding to the two erased lines of slope 1 in $(c_{u,v})$.
\qed

\begin{corollary}
\label{cor2}
{\em
The code $EBR(p,2,q,1)$ is MDS on lines of slope 0, 1 and $\infty$.
}
\end{corollary}

\begin{example}
\label{ex4}
{\em
Consider the code $EBR(7,2,q,1)$. The transpose transformation as
described in Lemma~\ref{l3} gives

\begin{eqnarray*}
\left(\begin{array}{ccccccc}
c_{0,0}&c_{0,1}&c_{0,2}&c_{0,3}&c_{0,4}&c_{0,5}&c_{0,6}\\
c_{1,0}&c_{1,1}&c_{1,2}&c_{1,3}&c_{1,4}&c_{1,5}&c_{1,6}\\
c_{2,0}&c_{2,1}&c_{2,2}&c_{2,3}&c_{2,4}&c_{2,5}&c_{2,6}\\
c_{3,0}&c_{3,1}&c_{3,2}&c_{3,3}&c_{3,4}&c_{3,5}&c_{3,6}\\
c_{4,0}&c_{4,1}&c_{4,2}&c_{4,3}&c_{4,4}&c_{4,5}&c_{4,6}\\
c_{5,0}&c_{5,1}&c_{5,2}&c_{5,3}&c_{5,4}&c_{5,5}&c_{5,6}\\
c_{6,0}&c_{6,1}&c_{6,2}&c_{6,3}&c_{6,4}&c_{6,5}&c_{6,6}\\
\end{array}\right)
&\rightarrow &
\left(\begin{array}{ccccccc}
c_{0,0}&c_{1,0}&c_{2,0}&c_{3,0}&c_{4,0}&c_{5,0}&c_{6,0}\\
c_{0,1}&c_{1,1}&c_{2,1}&c_{3,1}&c_{4,1}&c_{5,1}&c_{6,1}\\
c_{0,2}&c_{1,2}&c_{2,2}&c_{3,2}&c_{4,2}&c_{5,2}&c_{6,2}\\
c_{0,3}&c_{1,3}&c_{2,3}&c_{3,3}&c_{4,3}&c_{5,3}&c_{6,3}\\
c_{0,4}&c_{1,4}&c_{2,4}&c_{3,4}&c_{4,4}&c_{5,4}&c_{6,4}\\
c_{0,5}&c_{1,5}&c_{2,5}&c_{3,5}&c_{4,5}&c_{5,5}&c_{6,5}\\
c_{0,6}&c_{1,6}&c_{2,6}&c_{3,6}&c_{4,6}&c_{5,6}&c_{6,6}\\
\end{array}\right)
\end{eqnarray*}

We can see that columns become rows, rows become columns, and the
lines of slope~1 are preserved by this transformation, so the transposed
arrays are also in $EBR(7,2,q,1)$ and any pair of erased horizontal
lines can be recovered.

The second transformation in
Lemma~\ref{l3}, i.e., $b_{u,v}\eq c_{\lan -u\ran,\lan u+v\ran}$,
gives the following correspondence:

\begin{eqnarray*}
\left(\begin{array}{ccccccc}
c_{0,0}&c_{0,1}&{\bf c_{0,2}}&c_{0,3}&c_{0,4}&c_{0,5}&c_{0,6}\\
c_{1,0}&{\bf c_{1,1}}&c_{1,2}&c_{1,3}&c_{1,4}&c_{1,5}&c_{1,6}\\
{\bf c_{2,0}}&c_{2,1}&c_{2,2}&c_{2,3}&c_{2,4}&c_{2,5}&c_{2,6}\\
c_{3,0}&c_{3,1}&c_{3,2}&c_{3,3}&c_{3,4}&c_{3,5}&{\bf c_{3,6}}\\
c_{4,0}&c_{4,1}&c_{4,2}&c_{4,3}&c_{4,4}&{\bf c_{4,5}}&c_{4,6}\\
c_{5,0}&c_{5,1}&c_{5,2}&c_{5,3}&{\bf c_{5,4}}&c_{5,5}&c_{5,6}\\
c_{6,0}&c_{6,1}&c_{6,2}&{\bf c_{6,3}}&c_{6,4}&c_{6,5}&c_{6,6}\\
\end{array}\right)
&\rightarrow &
\left(\begin{array}{ccccccc}
c_{0,0}&c_{0,1}&{\bf c_{0,2}}&c_{0,3}&c_{0,4}&c_{0,5}&c_{0,6}\\
c_{6,1}&c_{6,2}&{\bf c_{6,3}}&c_{6,4}&c_{6,5}&c_{6,6}&c_{6,0}\\
c_{5,2}&c_{5,3}&{\bf c_{5,4}}&c_{5,5}&c_{5,6}&c_{5,0}&c_{5,1}\\
c_{4,3}&c_{4,4}&{\bf c_{4,5}}&c_{4,6}&c_{4,0}&c_{4,1}&c_{4,2}\\
c_{3,4}&c_{3,5}&{\bf c_{3,6}}&c_{3,0}&c_{3,1}&c_{3,2}&c_{3,3}\\
c_{2,5}&c_{2,6}&{\bf c_{2,0}}&c_{2,1}&c_{2,2}&c_{2,3}&c_{2,4}\\
c_{1,6}&c_{1,0}&{\bf c_{1,1}}&c_{1,2}&c_{1,3}&c_{1,4}&c_{1,5}\\
\end{array}\right)
\end{eqnarray*}

We can see that this transformation maps lines of slope 1 in
$(c_{u,v})$ into lines of slope $\infty$ in $(b_{u,v})$, lines of
slope 0 in $(c_{u,v})$ into lines of slope 0 in $(b_{u,v})$ and
lines of slope $\infty$ in
$(c_{u,v})$ into lines of slope 1 in $(b_{u,v})$. For example, the
line of slope~1 through $c_{2,0}$ (in bold) is mapped into the
third vertical line.
So, $(b_{u,v})\in EBR(7,2,q,1)$ and any pair of erased lines of
slope~1 can be recovered.
}
\end{example}

Consider next the case $r\eq 3$.

\begin{lemma}
\label{l4}
{\em
The code $EBR(p,3,q,1)$ can recover any three erased lines of
lines of slope $j$, where $0\leq j\leq 2$.
}
\end{lemma}

\noindent\pf
Assume that $(c_{u,v})$, $0\leq u,v\leq p-1$, is an array in
$EBR(p,3,q,1)$, and assume that three rows (lines of slope 0) have been
erased. Consider the
array  $(b_{u,v})$, $0\leq u,v\leq p-1$, such that
$b_{u,v}\eq c_{\lan 2v\ran,u}$. Then lines of slope
$\infty$ in $(c_{u,v})$ are mapped into lines of slope 0 in
$(b_{u,v})$ and lines of slope
0 in $(c_{u,v})$ are mapped into lines of slope $\infty$ in
$(b_{u,v})$, like in the case of the transpose transformation. Lines
of slope~2 in $(c_{u,v})$ are
mapped into lines of slope~1 in
$(b_{u,v})$. In effect, consider the line of slope~1 in $(b_{u,v})$
through entry $b_{u_0,0}$, where $0\leq u_0\leq p-1$. According to
Definition~\ref{def0}, this line
corresponds to the set $\{b_{\lan u_0-v\ran,v}\,:\,0\leq v\leq
p-1\}$, which is equal to the set
$\{c_{\lan 2v\ran,\lan u_0-v\ran}\,,\,0\leq v\leq p-1\}$. Since $\lan
2v\ran +\lan 2(u_0-v)\ran\eq \lan 2u_0\ran$, then this last set corresponds
to the line of slope~2 in $(c_{u,v})$ through entry $(\lan 2u_0\ran,0)$.

Similarly, proceeding as in the previous cases, it can be shown that lines of slope~1 in $(c_{u,v})$ are
mapped into lines of slope~2 in $(b_{u,v})$. Thus, $(b_{u,v})$
is in $EBR(p,3,q,1)$, hence, by Theorem~\ref{theo1}, it can correct
any three erased columns (which correspond to three erased rows in $(c_{u,v})$).

Next assume that three lines of slope 1 have been
erased in $(c_{u,v})$. We can consider the same transformation as in
Lemma~\ref{l3}, i.e., consider a $p\times p$ array $(b_{u,v})$ such that
$b_{u,v}\eq c_{\lan -u\ran,\lan u+v\ran}$
for $0\leq u,v\leq p-1$. As in Lemma~\ref{l3},
every line of slope $\infty$ in $(b_{u,v})$ corresponds to a line of
slope 1 in $(c_{u,v})$, each line of slope 0 in
$(b_{u,v})$ corresponds to a line of slope 0 in $(c_{u,v})$
and each line of slope 1 in
$(b_{u,v})$ corresponds to a line of slope $\infty$ in $(c_{u,v})$.
In addition, proceeding as in the previous cases, it can be shown
that each line of slope 2 in
$(b_{u,v})$ corresponds to a line of slope 2 in $(c_{u,v})$. Thus,
$(b_{u,v})\in EBR(p,3,q,1)$ and any 3 columns in $(b_{u,v})$, which
correspond to 3 lines of slope 1 in $(c_{u,v})$, can be corrected.

Finally, assume that three lines of slope 2 have been
erased in $(c_{u,v})$. Consider a $p\times p$ array $(b_{u,v})$ such that\\
$b_{u,v}\eq c_{\lan -2(u+v)\ran,\lan u+2v\ran}$
for $0\leq u,v\leq p-1$. Now, it can be shown that every line of
slope $\infty$ in
$(b_{u,v})$ corresponds to a line of
slope 2 in $(c_{u,v})$, every line of slope 0 in
$(b_{u,v})$ corresponds to a line of
slope 1 in $(c_{u,v})$, every line of slope 1 in
$(b_{u,v})$ corresponds to a line of
slope 0 in $(c_{u,v})$ and every line of slope 2 in
$(b_{u,v})$ corresponds to a line of
slope $\infty$ in $(c_{u,v})$. Thus,
$(b_{u,v})\in EBR(p,3,q,1)$ and any 3 columns in $(b_{u,v})$, which
correspond to 3 lines of slope 2 in $(c_{u,v})$, can be corrected.
\qed

\begin{corollary}
\label{cor3}
{\em
The code $EBR(p,3,q,1)$ is MDS on lines of slope 0, 1, 2 and $\infty$.
}
\end{corollary}

\begin{example}
\label{ex5}
{\em
Consider the code $EBR(7,3,q,1)$. The first transformation
in Lemma~\ref{l4}, i.e., $b_{u,v}\eq c_{\lan 2v\ran,u}$, gives

\begin{eqnarray*}
\left(\begin{array}{ccccccc}
c_{0,0}&c_{0,1}&{\bf c_{0,2}}&c_{0,3}&c_{0,4}&c_{0,5}&c_{0,6}\\
c_{1,0}&{\bf c_{1,1}}&c_{1,2}&c_{1,3}&c_{1,4}&c_{1,5}&c_{1,6}\\
{\bf c_{2,0}}&c_{2,1}&c_{2,2}&c_{2,3}&c_{2,4}&c_{2,5}&c_{2,6}\\
c_{3,0}&c_{3,1}&c_{3,2}&c_{3,3}&c_{3,4}&c_{3,5}&{\bf c_{3,6}}\\
c_{4,0}&c_{4,1}&c_{4,2}&c_{4,3}&c_{4,4}&{\bf c_{4,5}}&c_{4,6}\\
c_{5,0}&c_{5,1}&c_{5,2}&c_{5,3}&{\bf c_{5,4}}&c_{5,5}&c_{5,6}\\
c_{6,0}&c_{6,1}&c_{6,2}&{\bf c_{6,3}}&c_{6,4}&c_{6,5}&c_{6,6}\\
\end{array}\right)
&\rightarrow &
\left(\begin{array}{ccccccc}
c_{0,0}&{\bf c_{2,0}}&c_{4,0}&c_{6,0}&c_{1,0}&c_{3,0}&c_{5,0}\\
c_{0,1}&c_{2,1}&c_{4,1}&c_{6,1}&{\bf c_{1,1}}&c_{3,1}&c_{5,1}\\
{\bf c_{0,2}}&c_{2,2}&c_{4,2}&c_{6,2}&c_{1,2}&c_{3,2}&c_{5,2}\\
c_{0,3}&c_{2,3}&c_{4,3}&{\bf c_{6,3}}&c_{1,3}&c_{3,3}&c_{5,3}\\
c_{0,4}&c_{2,4}&c_{4,4}&c_{6,4}&c_{1,4}&c_{3,4}&{\bf c_{5,4}}\\
c_{0,5}&c_{2,5}&{\bf c_{4,5}}&c_{6,5}&c_{1,5}&c_{3,5}&c_{5,5}\\
c_{0,6}&c_{2,6}&c_{4,6}&c_{6,6}&c_{1,6}&{\bf c_{3,6}}&c_{5,6}\\
\end{array}\right)
\end{eqnarray*}

We can see that lines of slope $0$ in $(c_{u,v})$ become lines of
slope $\infty$ in $(b_{u,v})$, lines of slope $\infty$ in $(c_{u,v})$ become lines of
slope $0$ in $(b_{u,v})$, lines of slope 1 in $(c_{u,v})$ become lines of
slope 2 in $(b_{u,v})$ and lines of slope 2 in $(c_{u,v})$ become lines of
slope 1 in $(b_{u,v})$, so  $(b_{u,v})\in EBR(7,3,q,1)$ and any three
erased horizontal lines can be recovered. For example,
the line of slope~1 starting in $c_{2,0}$ (in bold) is mapped into the
line of slope~2 in the right array, also in bold.

The second transformation in
Lemma~\ref{l4}, i.e., $b_{u,v}\eq c_{\lan -u\ran,\lan u+v\ran}$, is
the same as the second transformation in Lemma~\ref{l3} and has been
illustrated in Example~\ref{ex4}. As in Example~\ref{ex4},
this transformation maps lines of slope 1 in
$(c_{u,v})$ into lines of slope $\infty$ in $(b_{u,v})$, lines of
slope 0 in $(c_{u,v})$ into lines of slope 0 in $(b_{u,v})$ and
lines of slope $\infty$ in
$(c_{u,v})$ into lines of slope 1 in $(b_{u,v})$. In addition, lines
of slope 2 in $(c_{u,v})$ are mapped into lines of slope 2 in
$(b_{u,v})$, hence, $(b_{u,v})\in EBR(7,3,q,1)$ and any three erased lines of
slope 1 can recovered.

Finally, the last transformation
in Lemma~\ref{l4}, i.e., $b_{u,v}\eq c_{\lan -2(u+v)\ran,\lan u+2v\ran}$,
gives

\begin{eqnarray*}
\left(\begin{array}{ccccccc}
c_{0,0}&{\bf c_{0,1}}&c_{0,2}&c_{0,3}&c_{0,4}&c_{0,5}&c_{0,6}\\
c_{1,0}&c_{1,1}&c_{1,2}&c_{1,3}&{\bf c_{1,4}}&c_{1,5}&c_{1,6}\\
{\bf c_{2,0}}&c_{2,1}&c_{2,2}&c_{2,3}&c_{2,4}&c_{2,5}&c_{2,6}\\
c_{3,0}&c_{3,1}&c_{3,2}&{\bf c_{3,3}}&c_{3,4}&c_{3,5}&c_{3,6}\\
c_{4,0}&c_{4,1}&c_{4,2}&c_{4,3}&c_{4,4}&c_{4,5}&{\bf c_{4,6}}\\
c_{5,0}&c_{5,1}&{\bf c_{5,2}}&c_{5,3}&c_{5,4}&c_{5,5}&c_{5,6}\\
c_{6,0}&c_{6,1}&c_{6,2}&c_{6,3}&c_{6,4}&{\bf c_{6,5}}&c_{6,6}\\
\end{array}\right)
&\rightarrow &
\left(\begin{array}{ccccccc}
c_{0,0}&{\bf c_{5,2}}&c_{3,4}&c_{1,6}&c_{6,1}&c_{4,3}&c_{2,5}\\
c_{5,1}&{\bf c_{3,3}}&c_{1,5}&c_{6,0}&c_{4,2}&c_{2,4}&c_{0,6}\\
c_{3,2}&{\bf c_{1,4}}&c_{6,6}&c_{4,1}&c_{2,3}&c_{0,5}&c_{5,0}\\
c_{1,3}&{\bf c_{6,5}}&c_{4,0}&c_{2,2}&c_{0,4}&c_{5,6}&c_{3,1}\\
c_{6,4}&{\bf c_{4,6}}&c_{2,1}&c_{0,3}&c_{5,5}&c_{3,0}&c_{1,2}\\
c_{4,5}&{\bf c_{2,0}}&c_{0,2}&c_{5,4}&c_{3,6}&c_{1,1}&c_{6,3}\\
c_{2,6}&{\bf c_{0,1}}&c_{5,3}&c_{3,5}&c_{1,0}&c_{6,2}&c_{4,4}\\
\end{array}\right)
\end{eqnarray*}

We can see that this transformation maps lines of slope 2 in
$(c_{u,v})$ into lines of slope $\infty$ in $(b_{u,v})$, lines of
slope 0 in $(c_{u,v})$ into lines of slope 1 in $(b_{u,v})$,
lines of slope $\infty$ in $(c_{u,v})$ into lines of slope 2 in
$(b_{u,v})$ and lines of slope 1 in $(c_{u,v})$ into lines of slope 0
in $(b_{u,v})$. For example, the line of slope~2 starting
in $c_{2,0}$ (in bold) is mapped into the second vertical line.

In particular, $(b_{u,v})\in EBR(7,3,q,1)$ and any three
erased lines of slope 2 can recovered.
}
\end{example}

We have seen that an $EBR(p,r,q,1)$ code is MDS on lines of
slope $j$, where $0\leq j\leq r-1$ and $1\leq r\leq 3$.
The next lemma shows that this is also the case for $p-2\leq r\leq p-1$.

\begin{lemma}
\label{l5}
{\em
The code $EBR(p,r,q,1)$ with $p-2\leq r\leq p-1$ is MDS on
lines of slope $j$, where $0\leq j\leq r-1$.
}
\end{lemma}

\noindent\pf
Consider an array $(c_{u,v})\in EBR(p,p-2,q,1)$.
Take $j$ such that $0\leq j\leq p-3$ and assume that $r$ lines
of slope $j$ have been erased.
Consider the
array $(b_{u,v})$, $0\leq u,v\leq p-1$, such that
$b_{u,v}\eq c_{\lan -ju+(3j+2)v\ran,\lan u+jv\ran}$.
Lines of slope $j$ in $(c_{u,v})$ are mapped into lines of slope $\infty$ in
$(b_{u,v})$. In effect, consider the line of slope $\infty$ in $(b_{u,v})$
through entry $b_{0,v_0}$, where\\ $0\leq v_0\leq p-1$. This line
corresponds to the set $\{b_{u,v_0}\,:\,0\leq u\leq p-1\}$, which is equal to the set\\
$\{c_{\lan -ju+(3j+2)v_0\ran,\lan u+jv_0\ran}\,,\,0\leq u\leq p-1\}$.
Since  $\lan -ju+(3j+2)v_0\ran+j\lan u+jv_0\ran\eq \lan
(j^2+2j+3)v_0\ran$, then, according to Definition~\ref{def0}, this last set corresponds
to the line of slope~$j$ in $(c_{u,v})$ through entry $(\lan (j^2+2j+3)v_0\ran,0)$.
Since\\ $\lan j^2+2j+3\ran\eq \lan(j+1)(j+2)\ran$ and $0\leq j\leq
p-3$, $\lan j^2+2j+3\ran\neq 0$, so to each choice of $v_0$
corresponds a unique line of slope $j$.

Proceeding similarly, we can show that lines of slope $p-1$ in $(c_{u,v})$ are
mapped into lines of slope $p-2$ in
$(b_{u,v})$ and that lines of slope $p-2$ in $(c_{u,v})$ are
mapped into lines of slope $p-1$ in $(b_{u,v})$.
Thus, a line of slope $i_0$ in $(c_{u,v})$, $i_0\neq j$, $i_0\eq\infty$ or $0\leq i_0\leq
r-1$, is mapped into a line of slope $i_1$ in $(b_{u,v})$,
$i_1\not\in\{\infty,p-2,p-1\}$. Since these lines have even parity,
$(b_{u,v})$ is in $EBR(p,p-2,q,1)$ and it can correct any $p-2$ erased columns,
which correspond to $p-2$ erased lines of slope $j$ in $(c_{u,v})$.

Next let $r\eq p-1$. Consider the
array $(b_{u,v})$, $0\leq u,v\leq p-1$, such that
$b_{u,v}\eq c_{\lan -ju+(j+1)v\ran,\lan u\ran}$. Proceeding as above,
we can verify that
lines of slope $j$ in $(c_{u,v})$ are mapped into lines of slope $\infty$ in
$(b_{u,v})$ and
that lines of slope $p-1$ in $(c_{u,v})$ are
mapped into lines of slope $p-1$ in
$(b_{u,v})$. Thus, a line of slope $i_0$, $i_0\neq p-1$, will be mapped
to a line of slope $i_1$, $i_1\neq p-1$. All these lines have even
parity, so $(b_{u,v})$ is in $EBR(p,p-2,q,1)$ and it can correct any
$p-1$ erased columns,
which correspond to $p-1$ erased lines of slope $j$ in $(c_{u,v})$.
\qed

\begin{example}
\label{ex6}
{\em
As in Example~\ref{ex5}, take $p\eq 7$ and consider the code
$EBR(7,5,q,1)$. If $j\eq 0$, the transformation
in Lemma~\ref{l5}, i.e., $b_{u,v}\eq c_{\lan -ju+(3j+2)v\ran,\lan
u+jv\ran}$, becomes $b_{u,v}\eq c_{\lan 2v\ran,u}$, giving the first
transformation illustrated in Example~\ref{ex5}. We have seen in this
example that the resulting array $(b_{u,v})$ is in $EBR(7,3,q,1)$ and
hence any tree rows can be corrected. In addition, we observe that
lines of slope~3 in $(c_{u,v})$ are transformed into lines of slope~3
in $(b_{u,v})$, and that lines of slope~4 in $(c_{u,v})$ are
transformed into lines of slope~4 in $(b_{u,v})$. Hence, if
$(c_{u,v})$ is in $EBR(7,5,q,1)$ also $(b_{u,v})$ is in
$EBR(7,5,q,1)$ and any 5 erased rows can be recovered.

Let us take next $j\eq 3$, then the transformation
$b_{u,v}\eq c_{\lan -ju+(3j+2)v\ran,\lan
u+jv\ran}$, becomes
$b_{u,v}\eq c_{\lan 4u+4v\ran,\lan u+3v\ran}$, giving

\begin{eqnarray*}
\left(\begin{array}{ccccccc}
c_{0,0}&c_{0,1}&c_{0,2}&c_{0,3}&c_{0,4}&c_{0,5}&c_{0,6}\\
c_{1,0}&c_{1,1}&c_{1,2}&c_{1,3}&c_{1,4}&c_{1,5}&c_{1,6}\\
c_{2,0}&c_{2,1}&c_{2,2}&c_{2,3}&c_{2,4}&c_{2,5}&c_{2,6}\\
c_{3,0}&c_{3,1}&c_{3,2}&c_{3,3}&c_{3,4}&c_{3,5}&c_{3,6}\\
c_{4,0}&c_{4,1}&c_{4,2}&c_{4,3}&c_{4,4}&c_{4,5}&c_{4,6}\\
c_{5,0}&c_{5,1}&c_{5,2}&c_{5,3}&c_{5,4}&c_{5,5}&c_{5,6}\\
c_{6,0}&c_{6,1}&c_{6,2}&c_{6,3}&c_{6,4}&c_{6,5}&c_{6,6}\\
\end{array}\right)
&\rightarrow &
\left(\begin{array}{ccccccc}
c_{0,0}&c_{4,3}&c_{1,6}&c_{5,2}&c_{2,5}&c_{6,1}&c_{3,4}\\
c_{4,1}&c_{1,4}&c_{5,0}&c_{2,3}&c_{6,6}&c_{3,2}&c_{0,5}\\
c_{1,2}&c_{5,5}&c_{2,1}&c_{6,4}&c_{3,0}&c_{0,3}&c_{4,6}\\
c_{5,3}&c_{2,6}&c_{6,2}&c_{3,5}&c_{0,1}&c_{4,4}&c_{1,0}\\
c_{2,4}&c_{6,0}&c_{3,3}&c_{0,6}&c_{4,2}&c_{1,5}&c_{5,1}\\
c_{6,5}&c_{3,1}&c_{0,4}&c_{4,0}&c_{1,3}&c_{5,6}&c_{2,2}\\
c_{3,6}&c_{0,2}&c_{4,5}&c_{1,1}&c_{5,4}&c_{2,0}&c_{6,3}\\
\end{array}\right)
\end{eqnarray*}

We can see that this transformation maps lines of slope 3 in
$(c_{u,v})$ into lines of slope $\infty$ in $(b_{u,v})$, lines of
slope $\infty$ in $(c_{u,v})$ into lines of slope 3 in $(b_{u,v})$,
lines of slope 0 in $(c_{u,v})$ into lines of slope 1 in
$(b_{u,v})$, lines of slope 1 in $(c_{u,v})$ into lines of slope 0 in
$(b_{u,v})$,
lines of slope 2 in $(c_{u,v})$ into lines of slope 4
in $(b_{u,v})$ and lines of slope 4 in $(c_{u,v})$ into lines of slope 2
in $(b_{u,v})$. In particular, $(b_{u,v})\in EBR(7,5,q,1)$ and any 5
erased columns, which correspond to 5 lines of slope 3 in
$(c_{u,v})$, can be recovered.

Similar transformations can be obtained for $j\eq 1$, 2 and 4.
}
\end{example}

Let us unify Theorem~\ref{theo1}, Corollaries~\ref{cor2}
and~\ref{cor3} and Lemma~\ref{l5}
in the following theorem:

\begin{theorem}
\label{theo2}
{\em
The code $EBR(p,r,q,1)$ with $1\leq r\leq 3$ or $p-2\leq r\leq p-1$ is MDS on
lines of slope $j$, where $j\eq \infty$ or $0\leq j\leq r-1$.
\qed
}
\end{theorem}

Theorem~\ref{theo2} gives five values of $r$ for which the code
$EBR(p,r,q,1)$ is MDS on lines of slope $j$, where $j\eq \infty$ or\\
$0\leq j\leq r-1$.
For $4\leq r\leq p-3$ we do not  think that this is the case, but the
problem is open.

\section{Puncturing EBR and EIP Codes}
\label{punctured}
A puncturing of a code of length $n$ in $s$ specified locations is a
code of length $n-s$ such that the $s$ specified
entries in each codeword of the original code are
deleted~\cite{ms}.
For $EBR(p,r,q,g(x))$ codes we will puncture
some rows as follows:

\begin{definition}
\label{defPEBR}
{\em
Consider an $EBR(p,r,q,g(x))$ code (resp., an $EIP(p,r,q,g(x))$ code)
and assume that $g(x)$ has degree $t$. A
punctured EBR (resp. EIP) code $PEBR(p,r,q,g(x))$ (resp.
$PEIP(p,r,q,g(x))$) consists of all $(p-t-1)\times p$  (resp.\\ $(p-t-1)\times (p+r)$)
arrays obtained by deleting the last $t+1$ rows of each array in
$EBR(p,r,q,g(x))$ (resp., in\\ $EIP(p,r,q,g(x))$).
\qed
}
\end{definition}

The next lemma is immediate.

\begin{lemma}
\label{l11}
{\em
The code $PEBR(p,r,q,g(x))$ as given by Definition~\ref{defPEBR} is
MDS (on columns), while the code $PEIP(p,r,q,g(x))$ is MDS if and
only if the code $EIP(p,r,q,g(x))$ is MDS.
}
\end{lemma}
\pf Simply observe that a column in $EBR(p,r,q,g(x))$ (resp., in
$EIP(p,r,q,g(x))$) is a zero column if and only if the corresponding
column in $PEBR(p,r,q,g(x))$ is a zero column (resp. in
$PEIP(p,r,q,g(x))$). In effect, using the notation of
Definition~\ref{defPEBR}, a vector in $\C(p,g(x)(1\xor x),q,d)$
is the zero vector if and only if the first $p-t-1$ entries of the
vector are zero, since encoding systematically such $p-t-1$ zero entries
in $\C(p,g(x)(1\xor x),q,d)$ gives the zero vector. \qed

Notice that in a $PEBR(p,r,q,g(x))$ or in a
$PEIP(p,r,q,g(x))$ code the vertical erasure-correction of a code is
lost. Let us examine next the simplest case of puncturing, that is,
when $g(x)\eq 1$.

\begin{example}
\label{ex10}
{\em
Consider a $PEBR(p,r,q,1)$ code, which, according to
Definition~\ref{defPEBR}, consists of all the $(p-1)\times p$ arrays
obtained by taking the first $p-1$ rows of each $p\times p$ array in
$EBR(p,r,q,1)$.

A $PEBR(p,r,q,1)$ code can be compared with a regular
$BR(p,r,q)$ code: they both consist of $(p-1)\times p$ arrays and can
recover up to $r$ erased columns. For example, the array in the left
below is in $PEBR(5,2,2,1)$, while the one in the right is in
$BR(5,2,2)$. The last row is not written. Notice that both arrays share the
same data array.

$$
\begin{array}{cc}
\begin{array}{|c|c|c||c|c|}
\hline
1&\rc{1}&1&\bc{0}&1\\\hline
\rc{1}&0&\bc{1}&1&1\\\hline
0&\bc{1}&1&0&\rc{0}\\\hline
\bc{0}&1&1&\rc{0}&0\\\hline\hline
0&1&\rc{0}&1&\bc{0}\\\hline
\end{array}
&
\begin{array}{|c|c|c||c|c|}
\hline
1&1&1&0&1\\\hline
1&0&1&1&1\\\hline
0&1&1&1&1\\\hline
0&1&1&1&1\\\hline\hline
0&0&0&0&0\\\hline
\end{array}
\end{array}
$$

However, the two codes $PEBR(p,r,q,1)$ and $BR(p,r,q)$ are not
equivalent. The code $PEBR(p,r,q,1)$ can correct up to $r$ lines of
slope $j$ for $1\leq j\leq r-1$ if $2\leq r\leq 3$ or $p-2\leq r\leq
p-1$. For example, consider the array in the left above in
$PEBR(5,2,2,1)$. The two lines of slope~1 colored in red and in blue
can be recovered. In effect, assuming that the array is in
$EBR(5,2,2,1)$ with the last row corresponding to erasures, applying
the transformation $b_{u,v}\eq c_{\lan -u\ran,\lan u+v\ran}$ for $0\leq
u,v\leq 4$ of Lemma~\ref{l3}, the lines of slope 1 become columns and
the resulting array is in $EBR(5,2,2,1)$. This transformed array has
two erased columns, while the remaining columns have one erasure
each and so it is recoverable by Theorem~\ref{theo1}. Once such array
is recovered, by applying
the inverse transformation to it, the original array in $EBR(5,2,2,1)$ is
obtained, and by deleting the last row, so is the array in $PEBR(5,2,2,1)$.
Codes $BR(p,r,q)$ on the other hand cannot recover two or more erased
lines that are not vertical, as shown in Example~\ref{ex0}.
}
\end{example}

The construction of a $PEIP(p,r,2,1)$ code is also 
given in~\cite{hscl2}. In Theorem~5 of this reference, it is proven that the code
$PEIP(p,r,2,1)$ is MDS when 2 is primitive in $GF(p)$, $p>5$ and
$r\leq 5$ using the Erd\"{o}s-Heilbronn conjecture. However, this
result is a special case of the combination of Corollary~\ref{cor00},
Lemma~\ref{l11} and Theorem~2.6 in~\cite{bbv}.

\begin{example}
\label{ex11}
{\em
Consider $EBR(7,3,2,1\xor x\xor x^3)$ as in Example~\ref{ex8}. Then,
$PEBR(7,3,2,1\xor x\xor x^3)$ consists of
all the $3\times 7$ arrays obtained by taking the
first three rows of each $7\times 7$ array in $EBR(7,3,2,1\xor x\xor x^3)$.

Taking the first 3 rows of the two arrays in
$EBR(7,3,2,1\xor x\xor x^3)$ of Example~\ref{ex8}, we obtain:

$$
\begin{array}{cc}
\begin{array}{|c|c|c|c||c|c|c|}
\hline
1&0&1&0&1&0&1\\
\hline
1&1&1&0&0&0&1\\
\hline
0&1&1&0&0&1&1\\
\hline
\end{array}
&
\begin{array}{|c|c|c|c||c|c|c|}
\hline
0&0&0&0&0&1&1\\
\hline
0&0&0&1&1&0&0\\
\hline
0&0&0&1&0&1&0\\
\hline
\end{array}
\end{array}
$$

Any three columns in the code can be recovered, so, in particular,
this code is a $[7,4,4]$ MDS code over vectors of length 3 (the
columns), like a RS code over $GF(8)$. In fact, it can be
shown that this code is equivalent to a $[7,4,4]$ RS code over $GF(8)$. In
effect, if we permute the last two rows of the arrays above, we
obtain

$$
\begin{array}{cc}
\begin{array}{|c|c|c|c||c|c|c|}
\hline
1&0&1&0&1&0&1\\
\hline
0&1&1&0&0&1&1\\
\hline
1&1&1&0&0&0&1\\
\hline
\end{array}
&
\begin{array}{|c|c|c|c||c|c|c|}
\hline
0&0&0&0&0&1&1\\
\hline
0&0&0&1&0&1&0\\
\hline
0&0&0&1&1&0&0\\
\hline
\end{array}
\end{array}
$$

Assuming that each column is a symbol in the finite field $GF(8)$
with $\beta$ a zero of the primitive polynomial $1\xor x\xor x^3$,
the first array above corresponds to the polynomial in $GF(8)$
$f_0(X)\eq\beta^6\xor\beta^4X\xor\beta^5X^2+ X^4\xor\beta X^5\xor\beta^5X^6$, while the
second one corresponds to
$f_1(X)\eq\beta^4X^3\xor\beta^2X^4\xor\beta^3X^5+ X^6$.
We can verify that $f_i(\beta^{-j})\eq 0$ for $0\leq i\leq 1$ and $0\leq
j\leq 2$, i.e., both codewords are in the RS code with generator
polynomial $(1\xor x)(\beta^{-1}\xor x)(\beta^{-2}\xor x)$. It can also be
verified more in general that $PEBR(7,r,2,1\xor x\xor x^3)$ with rows 1 and 2
permuted corresponds to a RS code over $GF(8)$ defined by the
primitive polynomial $1\xor x\xor x^3$ with generator polynomial $g(x)\eq
(1\xor x)\prod_{i=1}^{r-1}(\beta^{-i}\xor x)$. In the example above, $r\eq
3$.

On the other hand, if $\beta$ is a zero of the primitive polynomial
$1\xor x^2\xor x^3$, then, applying the permutation $(1\;2\;0)$ to the three
rows of the arrays in
$PEBR(7,r,2,1\xor x\xor x^3)$ corresponds to a RS code over $GF(8)$ with
generator polynomial $g(x)\eq
(1\xor x)\prod_{i=1}^{r-1}(\beta^{i}\xor x)$. The two original arrays with
this permutation are

$$
\begin{array}{cc}
\begin{array}{|c|c|c|c||c|c|c|}
\hline
1&1&1&0&0&0&1\\
\hline
0&1&1&0&0&1&1\\
\hline
1&0&1&0&1&0&1\\
\hline
\end{array}
&
\begin{array}{|c|c|c|c||c|c|c|}
\hline
0&0&0&1&1&0&0\\
\hline
0&0&0&1&0&1&0\\
\hline
0&0&0&0&0&1&1\\
\hline
\end{array}
\end{array}
$$
}
\end{example}

\begin{example}
\label{ex12}
{\em
Generalizing Example~\ref{ex11}, consider a Mersenne prime $p\eq
2^b-1$, where $b$ is also prime and $b\geq 3$. The first four such
Mersenne primes are
$7\eq 2^3-1$, $31\eq 2^5-1$, $127\eq 2^7-1$ and
$8191\eq 2^{13}-1$.

Consider a RS code over $GF(2^b)$, let $\beta$ be a primitive element
in $GF(2^b)$ and let $h(x)$ be a cyclotomic polynomial~\cite{ms} such
that $h(\beta)\eq 0$. Let $g(x)\eq (1\xor x^p)/(h(x)(1\xor x)$. Then,
the code $PEBR(p,r,2,g(x))$ is an MDS code on columns consisting of
$b\times p$ arrays. We believe that a permutation of the rows of such
arrays gives a code that is equivalent to a RS code over $GF(p+1)$ as
we showed for the case $p\eq 7$ in Example~\ref{ex11}, but we cannot
find a proof.
}
\end{example}



\section{Conclusions}
\label{conclusions}
We have expanded codes like Blaum-Roth codes and generalized
EVENODD codes to array codes such that each column has a certain
erasure-correcting capability. We have shown the connection of the
new codes to traditional array codes. We have presented efficient encoding,
decoding and updating algorithms. We have observed that the new codes
can recover erased lines of different slopes. We have also showed a
method for puncturing the codes such that the resulting arrays
constitute an MDS code.

\end{document}